\documentclass[journal=jpcl,manuscript=article]{achemso}

\usepackage[version=3]{mhchem} % Formula subscripts using \ce{}
\usepackage[labelfont=bf]{caption}
\usepackage[T1]{fontenc}
\usepackage[utf8]{inputenc}
\usepackage{multirow}

\newcommand{\comment}[1]{} % Command for commenting a block of text

\author{Miguel Riopedre}
\altaffiliation{These authors contributed equally to this work and are allowed to change the publication order to list them as first in their CVs}
\author{Denys Biriukov}
\altaffiliation{These authors contributed equally to this work and are allowed to change the publication order to list them as first in their CVs}
\author{Martin Dra\v{c}\'{i}nsk\'{y}}
\author{Hector Martinez-Seara}
\email{hseara@gmail.com}
\affiliation{Institute of Organic Chemistry and Biochemistry, Czech Academy of Sciences, Flemingovo nám. 2, 16610 Prague 6, Czech Republic}

\title{Hyaluronan-Arginine Enhanced Interaction Emerges from Distinctive Molecular Signature with Improved Electrostatics and Side-Chain Specificity}

\begin{document}

\clearpage

% \maketitle

\begin{abstract}
Hyaluronan, a sugar polymer found outside the plasma membrane, is a critical component of the extracellular matrix (ECM) scaffold. Hyaluronan's length and moderate negative charge aid in ECM entanglement and solubility. This allows it to effectively foster gel-like conditions, which are critical for supporting and protecting cells in various tissues. As a recognized ligand for multiple proteins, understanding its interaction with various amino acids helps elucidate its biological functions. This study employs NMR and molecular dynamics simulations (MD) to explore a potential general molecular mechanism for hyaluronan-protein interactions. We examine the binding of short arginine, lysine, or glycine oligopeptides to hyaluronan polysaccharides, using chemical shift perturbation (CSP) to detect changes in hyaluronan's 1H NMR spectra upon peptide titration. CSP helps to identify concrete binding regions and offers insights into the relative interaction strength of the peptides. We also use nuclear Overhauser effect (NOE) measurements to identify a potentially stable hyaluronan-peptide complex. However, the lack of NOE signals does not support the formation of strong and defined complexes. The combination of a positive CSP signal and negative NOE suggests a dynamic binding mode without stable complex formation, which MD further supports. Arginine, a positively charged amino acid, exhibits the strongest binding to the negatively charged hyaluronan, with the similarly charged lysine coming in distant second, highlighting the significance of electrostatics and side-chain specificity in these interactions. Hyaluronan's carboxyl and amide groups drive its interaction with the peptides, further stabilized by hydrogen bonds. Our findings elucidate hyaluronan-protein interaction patterns, typically involving arginine, and highlight arginine's crucial role in hyaluronan recognition by proteins.

\end{abstract}

\section{Introduction}

Hyaluronan, or hyaluronic acid (HA), is a major component of the sugar-enriched layer above the plasma membrane called the extracellular matrix, found in many mammalian cells \cite{Tarbell2016}. It is a natural polyelectrolyte constituted entirely by a repeating disaccharide of D-glucuronic acid (GlcUA, or GCU) and N-acetyl-D-glucosamine (GlcNAc, or NAG) ([-$\beta$(1,4)-GlcUA-$\beta$(1,3)-GlcNAc-]\textsubscript{n}) \cite{Toole2004}. HA can reach thousands of kDa molecular weights and is ubiquitously present in connective tissues, defining their elasticity and permeability \cite{Cowman2015}. Until recently, there were open debates about whether HA has only a passive protective function or actively participates in cell--cell communication, protein recognition, and disease development \cite{Fraser1997}. Current studies indicate that secretion, concentration, and weight of HA polysaccharide chains correlate with wound healing \cite{Aya2014}, arthritis development \cite{Zhang2015}, and even cancer resistance \cite{Tian2013}. Unfortunately, the complex and dynamic extracellular matrix structure entangles HA experimental and theoretical investigations in its natural environment. This complicates determining its possible modulating role in the structural organization of the extracellular matrix and governing cell membrane--protein interactions.

Despite many known HA-binding proteins \cite{Day2002}, the overall picture of their interactions remains elusive and mostly unresolved. For instance, HA binding sites of its primary transmembrane receptor, glycoprotein CD44, were characterized with sufficient accuracy and resolution only recently \cite{Vuorio2017}. One of the CD44 residues identified as crucial in CD44--HA interactions is R41, i.e., a positively charged arginine. Other arginine residues on the same protein face stabilize some binding modes. This agrees with intuitively accessible electrostatic interactions expected for negatively charged HA \cite{Sterling2021}. However, there is no apparent evidence whether electrostatics is the only (if any) driving force in HA--protein recognition.

By combining NMR measurements with molecular dynamics (MD) simulations, we demonstrate that HA exhibits a significantly stronger preference for arginine-containing peptide sequences. These interactions are not only promoted by electrostatics but are also enhanced by the specificity of the arginine side chain. The collected molecular information potentially offers a simple recipe for regulating HA interactions by synthesizing specific drug ligands or mutating existing proteins.

\section{Methods}

\subsection{Reagents}

The hyaluronan octamer (HA8\textsuperscript{AN}, from now on just HA8) and larger polysaccharide (HA\textsubscript{long}) with a molecular weight of 15-30~kDa were purchased as a powder from Contipro S. A. The fluorenylmethoxycarbonyl(FMOC)-amino acids were purchased from Iris Biotech. The tetrapeptides (R4, K4, and G4) were synthesized and capped (N-amidated and acetylated at C- and N-termini, respectively) \textit{in house} following a standard solid-phase peptide synthesis procedure \cite{Hansen2015}. The purification was performed in a water/MeOH gradient using high-performance liquid chromatography.

\subsection{NMR Measurements}

\subsubsection{Chemical shift perturbation}

The chemical shift perturbation method \cite{Williamson2013} was employed to assess possible interactions between hyaluronan and the studied tetrapeptides. The technique is based on the analysis of changes in the chemical shift of one substance upon the titration with another, which indicates possible interactions between them.

We prepared several solutions with varying amounts of HA8 while the concentration of hydrated peptides (R4, K4, G4) was kept constant (1.62~mM). The final peptide/HA8 molar ratios were 1:1, 1.5:1, and 2:1 for R4--H8 and K4--HA8 solutions, and 1:1 and 2:1 for G4--HA8 solutions. We also prepared solutions with HA\textsubscript{long} (with molar peptide/HA\textsubscript{long} ratios of 1:2 and 1:4) to verify that the observed trends remain the same irrespective of the molecular weight of the polysaccharides. To rule out the possibility of a concentration effect being the cause of the changes in the spectra, solutions of HA\textsubscript{long} with the same mass/volume concentrations as in the peptide--HA8 mixtures were measured in the absence of peptides.

All the solutions were prepared in D\textsubscript{2}O, and their \textsuperscript{1}H spectra were collected. The APT~\textsuperscript{13}C, HSQC, HMBC, and COSY spectra were additionally acquired for assignment purposes when necessary. The NMR spectra were recorded on a 400~MHz Bruker AVANCE III spectrometer (\textsuperscript{1}H at 401~MHz, \textsuperscript{13}C at 101~MHz) equipped with a liquid-nitrogen cryoprobe. The spectra were referenced to the solvent signal $\delta$(\textsuperscript{1}H) = 4.79~ppm (D\textsubscript{2}O). Every spectrum was obtained at 295~K. The resulting spectra were analyzed using the Mnova software \cite{Mestrelabresearch}. The position of the signals was selected using the Mnova tool "select peaks" to choose the position of the maximum unequivocally.

\subsubsection{Water suppression experiments}

Water suppression experiments were carried out for pure solutions of hydrated K4, R4, HA8, and their 1:1 mixtures (i.e., K4--HA8, R4--HA8). These experiments are required to identify labile protons like those of amide nitrogens, which are invisible in typical NMR experiments with deuterated water due to hydrogen-deuterium exchange. The peptide and HA8 concentrations were 2.44~mM in all cases, and the solutions were prepared in a 90:10 mixture of H\textsubscript{2}O/D\textsubscript{2}O. \textsuperscript{1}H, \textsuperscript{1}H-\textsuperscript{15}N HSQC, HMBC, and NOESY spectra were recorded at $\approx$ 295 K on a Bruker AVANCE III 600 MHz spectrometer (\textsuperscript{1}H at 600.1~MHz, \textsuperscript{13}C at 150.9~MHz, \textsuperscript{15}N at 60.8~MHz) and/or Bruker AVANCE III 500~MHz spectrometer (\textsuperscript{1}H at 500.0~MHz, \textsuperscript{13}C at 125.7~MHz, \textsuperscript{15}N at 50.7~MHz). The proton spectra measured in the H\textsubscript{2}O--D\textsubscript{2}O mixture were acquired with water signal suppression using excitation sculpting with gradients \cite{Hwang1995}. The signal assignment is based on homo- and heteronuclear correlation experiments, namely COSY, HSQC, HMBC, and NOESY.

\subsection{Molecular Dynamics}

\subsubsection{Computational models}

The hyaluronan octamer (HA8) was built in CHARMM-GUI \cite{Lee2015}, and the tetrapeptides (K4, R4, and G4) were drawn in Avogadro \cite{Hanwell2012}. We also modeled RKRK, A4, and P4 amino acid sequences for additional comparison. The C- and N-termini of each tetrapeptide were capped by N-amide and acetyl groups, respectively, to match the experimentally synthesized moieties. CHARMM-GUI was used to generate model topologies for both HA8 and tetrapeptides.

We employed two MD force fields in this work. The CHARMM36m force field is an all-atom generic biological force field containing parameters for all biomolecules, including proteins and polysaccharides \cite{Huang2016}. The prosECCo75 force field is a refined version of CHARMM36m that reintroduces the electronic polarization to the originally nonpolarizable force field via charge scaling \cite{Leontyev2011}. The only difference between these models is the atomic charges of ions and charged groups, i.e., peptide ammonium and guanidinium groups in side chains and hyaluronan carboxyl groups, see Table~\ref{tbl:force_field_charges}. The partial atomic charges of these ionic groups were scaled down so that the total charge of each group was 75\% of its original value. The charge scaling reflects the electronic polarization upon ionic solvation \cite{Nencini2022} and has been shown to improve the quality of MD simulations over a range of biological systems \cite{Duboue-Dijon2020}.

\subsubsection{Simulation protocol}

Each simulated system consisted of one hyaluronan octamer and one tetrapeptide. A 5~nm cubic box with both solutes initially separated from each other was solvated in CHARMM36m-compatible TIP3P water \cite{Jorgensen1983}. For neutral peptides (i.e., G4, A4, and P4), the system was neutralized by four potassium ions. All simulations were performed in the \emph{NPT} ensemble. The temperature of 300~K was maintained employing the Nose-Hoover thermostat with a coupling time of 1~ps \cite{Nose1984, Hoover1985}. The pressure of 1~bar was controlled via the Parrinello-Rahman barostat with a coupling time of 5~ps \cite{Parrinello1998}. The cut-off for electrostatic interactions was set to 1.2~nm, while the particle mesh Ewald method was used to calculate the long-range contribution \cite{Essmann1998}. The Lennard-Jones interactions were smoothly turned off between 1 to 1.2~nm using the force-based switching function \cite{Steinbach1994}. The water geometry was constrained using the SETTLE algorithm \cite{Miyamoto1992}. All other covalent bonds involving hydrogens were constrained using the LINCS algorithm \cite{Hess1997, Hess*2007}. All simulations were 1~$\mu$s long: the first 100~ns were treated as equilibration, and the last 900~ns were used for the analysis. All simulations were performed in Gromacs software, version 2021.1\cite{Abraham2015}. The MD trajectories were analyzed using \textit{in house} python scripts and post-processing inbox tools in Gromacs.

\section*{Results and Discussion}

\begin{figure}[htb!]
    \centering
    \includegraphics[width=\textwidth]{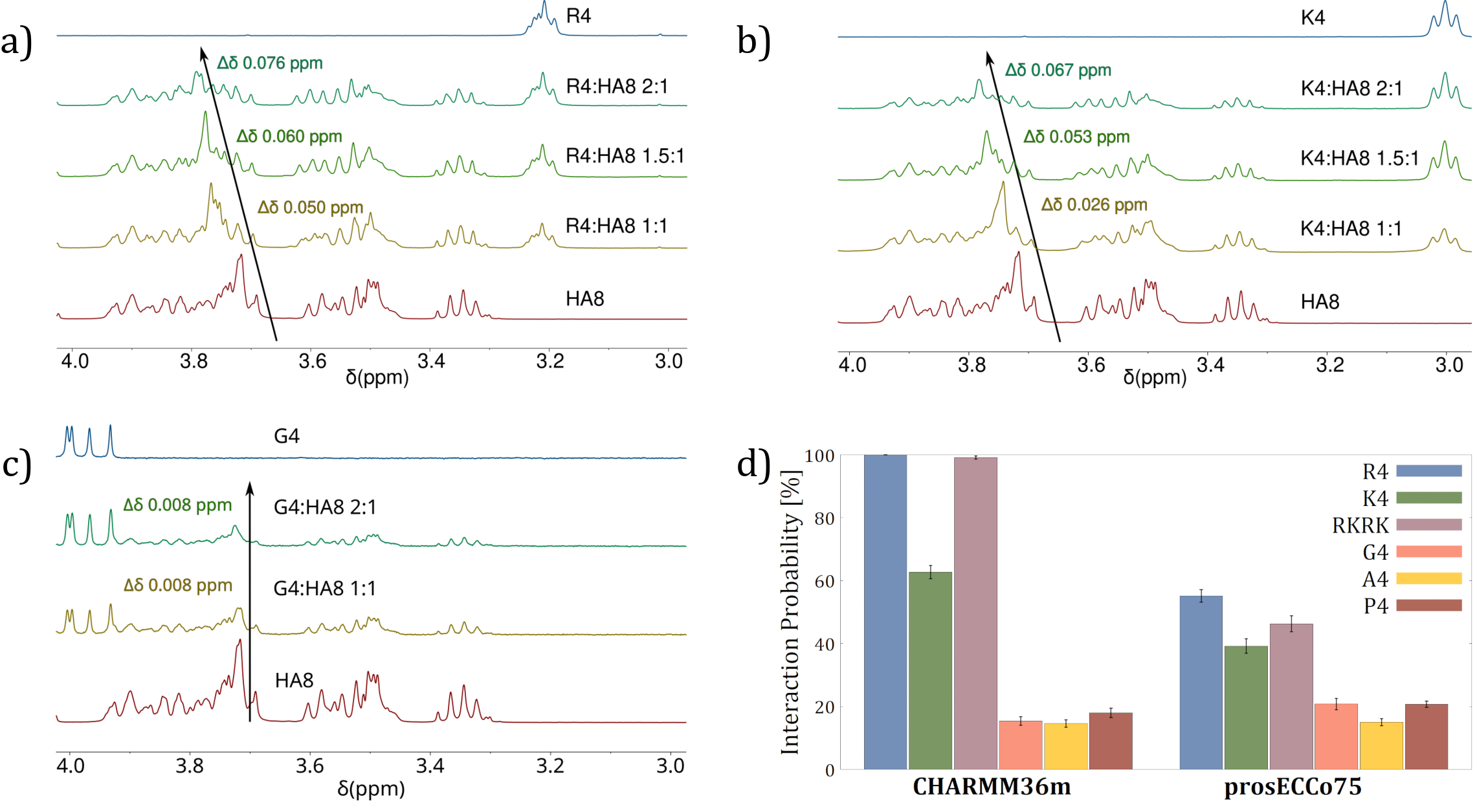}
    \caption{The \textsuperscript{1}H NMR spectra measured on hyaluronan octamer (HA8) solutions with a) R4, b) K4, and c) G4 tetrapeptides. The regions with the largest chemical shift perturbations are shown. The black arrows visually guide through the displacement (or its absence, like in the case of G4) of the chemical shifts upon adding a tetrapeptide to HA8 solution. The corresponding quantified shifts in the position of the signals are also given and colored following the spectra color scheme. d) The interaction probability of HA8 with modeled tetrapeptides calculated from molecular dynamics simulations with two different force field parameterizations. The probability is calculated as the fraction of simulation frames when any atoms of HA8 and tetrapeptide are closer than 0.35 nm.}
    \label{fgr:NMR_MD_summary}
\end{figure}

In this work, we investigate the binding affinity of oligopeptides to hyaluronan using a combination of NMR and MD methods. Hyaluronan has previously been suggested to preferentially interact with positively charged amino acids, as expected from its overall negative charge \cite{Vuorio2017,Chytil2016,Jugl2020,Ceely2023}. The comparison of collected \textsuperscript{1}H NMR spectra, Figures~\ref{fgr:NMR_MD_summary}a-c, indicates specific preferences for positively charged peptide sequences. This trend is reflected in notable concentration-dependent changes in chemical shifts upon adding R4 and K4 tetrapeptides to the solution with hyaluronan octamer. At the same time, NMR measurements on G4--HA8 mixtures show negligible alteration of the spectra at every concentration measured.

The MD data perfectly resemble the conclusions from interpreting the NMR results, Figure~\ref{fgr:NMR_MD_summary}d. We report the probability of tetrapeptides to interact with hyaluronan from corresponding simulations of tetrapeptide--HA8 solutions. Positively charged tetrapeptides notably prevail in binding affinity over neutrally charged, as seen from MD simulations with the two employed force field parameterizations. Both models eventually provide practically the same qualitative picture. The only significant difference is an even higher interaction probability for arginine- and lysine-containing tetrapeptides with the CHARMM36m model. This is not surprising because the CHARMM36m model is known to overestimate electrostatic interactions \cite{Duboue-Dijon2020} when no NBFIX parameters are available for a given charge pair. This is, unfortunately, the case for hyaluronan--peptide binding. Once electrostatic interactions are appropriately screened, as introduced in the prosECCo75 model, the interaction probability for positively charged tetrapeptides decreases but remains notably higher than for charge-neutral tetrapeptides.

NMR and MD data also point to a considerable difference between arginine and lysine in their affinity for hyaluronan, Figures~\ref{fgr:NMR_MD_summary}a-b,d. This discrepancy is nicely seen when comparing the NMR data for peptide--hyaluronan solutions with 1:1 molar ratios or the corresponding MD results. Moreover, the MD simulations with the CHARMM36m model suggest that tetraarginine is irreversibly bound to hyaluronan, which is not true for tetralysine. The results for the additionally modeled RKRK tetramer essentially show that the presence of two arginines increases the interaction probability to be almost as high as for tetraarginine, including introducing the irreversible character of interactions in CHARMM36m simulations. This observation also denotes a leading role of arginine in binding hyaluronan. At the same time, NMR data suggest a somewhat similar interaction mechanism for both R4 and K4 as given by qualitatively comparable changes in the NMR spectra of the corresponding peptide--hyaluronan solutions. We do not observe any remarkable difference in relatively weak interactions of A4, G4, and P4 (all neutral tetrapeptides) in either set of MD simulations. These data suggest that the strength of hyaluronan--peptide interaction is dictated mainly by electrostatics and only further governed by the side-chain specificity of basic amino acids.

The \textsuperscript{1}H NMR spectra measured on larger HA (HA\textsubscript{long}) have similar features as those measured on HA8 solutions, see Figures~\ref{fgr:NMR_R4_H8_1H}, \ref{fgr:NMR_K4_H8_1H}, \ref{fgr:NMR_G4_H8_1H} in the Supporting Information, where all the full spectra are summarized and compared. Upon adding positively charged tetramers, we observe the same signals affected and similar displacement magnitudes consistent with the measurements on tetrapeptide--HA8 mixtures. These results indicate that the binding affinities and interaction patterns are likely independent of polysaccharide molecular weight and peptide--hyaluronan molar ratio. We also confirmed that the chemical shift perturbations are caused by hyaluronan--peptide interactions, not by the change in hyaluronan concentration, see Figure~\ref{fgr:NMR_long_HA_conc}.

\begin{figure}[htb!]
    \centering
    \includegraphics[width=0.9\textwidth]{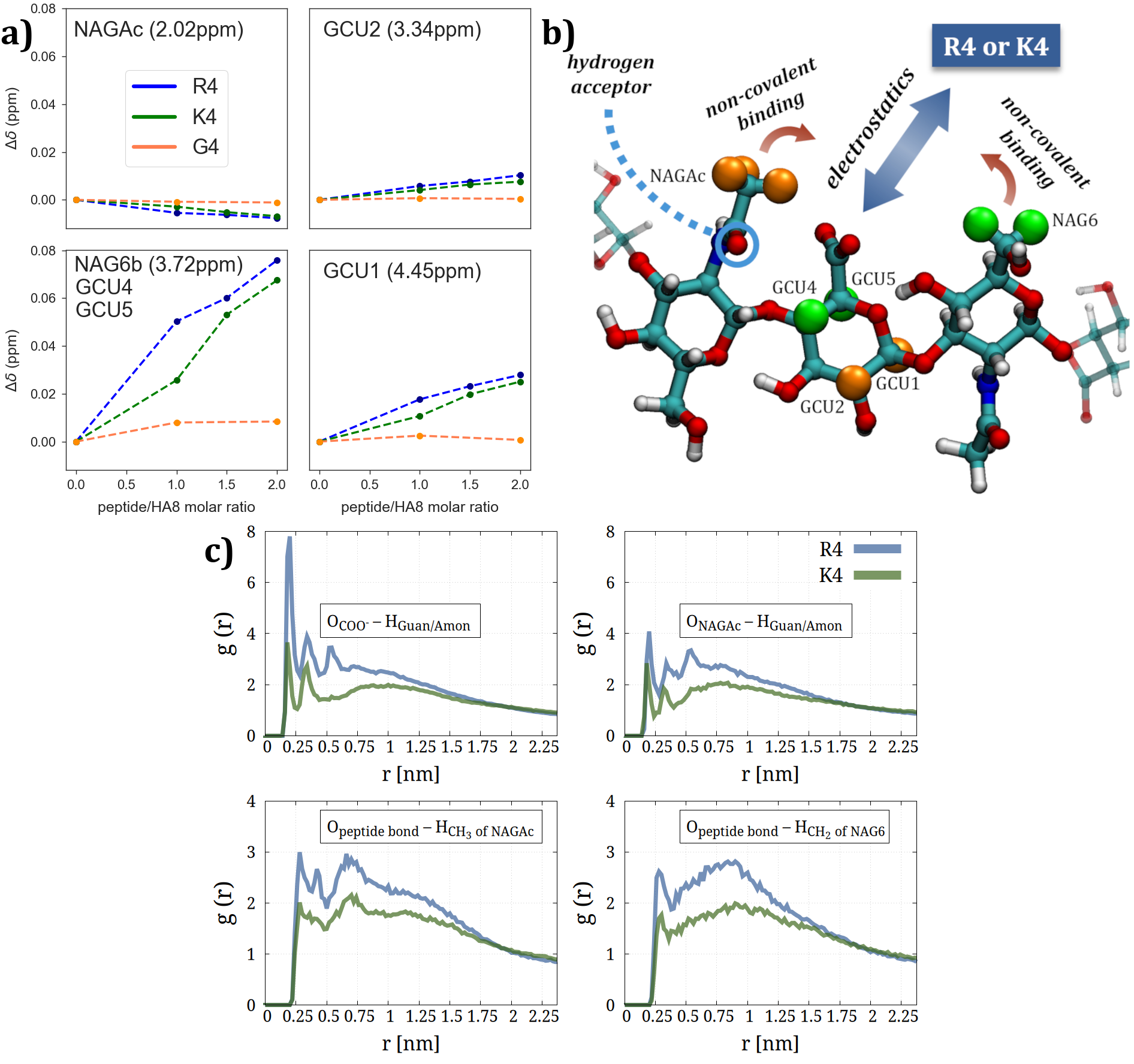}
    \caption{a) The summary of chemical shift perturbations ($\Delta\delta$) of selected HA8 peaks from \textsuperscript{1}H NMR spectra measurements after the addition of R4 (blue), K4 (green), or G4 (orange) tetrapeptide to HA8 solution. The data are presented as the difference between the position of the signal maximum in each mixture with respect to the reference, which is the position of the signal in pure HA8 solution (given in brackets). Each signal is assigned to certain HA8 proton(s) as shown in the Supporting Information, Figure~\ref{fgr:NMR_assign_ha}. b) The suggested molecular pattern for hyaluronan interactions with positively charged tetrapeptides. Enlarged colored spheres are hydrogens with the largest chemical shift perturbations measured by NMR. The green enlarged protons correspond to the largest displacement in the NMR signal at 3.72~ppm, and the orange enlarged spheres correspond to the rest of the identified chemical shift perturbations. c) The radial distribution functions for various atom groups of hyaluronan and tetrapeptides from molecular dynamics simulations with the prosECCo75 model.}
    \label{fgr:NMR_shifts_MD_data}
\end{figure}

Figure~\ref{fgr:NMR_shifts_MD_data}a show the concentration-dependent changes in the chemical shifts for hyaluronan--tetrapeptide mixtures of different molar ratios. The most significant perturbations (the order of 8$\times10^{-2}$~ppm) for R4--HA8 and K4--HA8 solutions are observed for the peak at 3.72~ppm. This peak corresponds to either GCU4, GCU5, or one of NAG6 protons. The complete signal assignment is provided in Figure~\ref{fgr:NMR_assign_ha}. These three protons are indistinguishable by NMR measurements alone, and it is impossible to tell if all or only some of them are affected by the presence of tetrapeptides. Interestingly, all three protons are close to hyaluronan carboxyl groups, see Figure~\ref{fgr:NMR_shifts_MD_data}b for a schematic binding pattern. This pattern suggests electrostatic interactions between the carboxyl groups and positively charged side chains of the tetrapeptides, which are stabilized by interactions with neighboring atoms or/and cause perturbations of neighboring protons. The same logic applies to changes in the signal at 2.02~ppm assigned to one of the acetamide (NAGAc) protons, also closely located to carboxyl groups in hyaluronan polymers.

The anticipated non-bonded interactions can be extracted from MD simulations as radial distribution functions (RDFs) shown in Figure~\ref{fgr:NMR_shifts_MD_data}c. One can see that carboxyl and amide oxygens of hyaluronan (O\textsubscript{COO\textsuperscript{--}} and O\textsubscript{NAGAc}, respectively) interact with side-chain cation hydrogens of basic amino acids (H\textsubscript{Guan} and H\textsubscript{Amin}), while methyl and methylene hydrogens of hyaluronan (H\textsubscript{CH\textsubscript{3} of NAGAc} and H\textsubscript{CH\textsubscript{2} of NAG6}) indeed mildly interact with amide oxygens of tetrapeptides (O\textsubscript{peptide bond}). Moreover, in the case of R4, hyaluronan--tetrapeptide binding can be stabilized by weak $\pi$-interactions of the guanidinium cation (the arginine side chain) with methyl/methylene hydrogens of hyaluronan, see Figure~\ref{fgr:MD_RDFs_pi}. All these MD results agree with the interpretation of the NMR data.

The displacements of signals at 3.34~ppm and 4.45~ppm correspond to chemical shift perturbations of GCU1 and GCU2 protons, respectively. Besides direct proximity contact, these changes might also indicate the possible bending of hyaluronan polymer upon interacting with positively charged peptides to optimize their interaction since both protons are close to the GlcUA-$\beta$(1,3)-GlcNAc bond. Noticeably, the chemical shifts of previously mentioned GCU4 and GCU5 hydrogens also can be altered by changes around the GlcNAc-$\beta$(1,4)-GlcUA bond. The possible reduction of the hyaluronan rigidity in the presence of R4 and K4 can induce or be induced by the reorientation of acetamide and hydroxymethyl groups. Most force fields, including CHARMM-based models, incorporate quite rigid glycosidic linkage for hyaluronan polymers \cite{Taweechat2020}. Therefore, we do not observe any perceptible bending in our MD simulations. Nevertheless, for short HA fragments, this is expected.

\begin{figure}[htb!]
    \centering
    \includegraphics[width=0.65\textwidth]{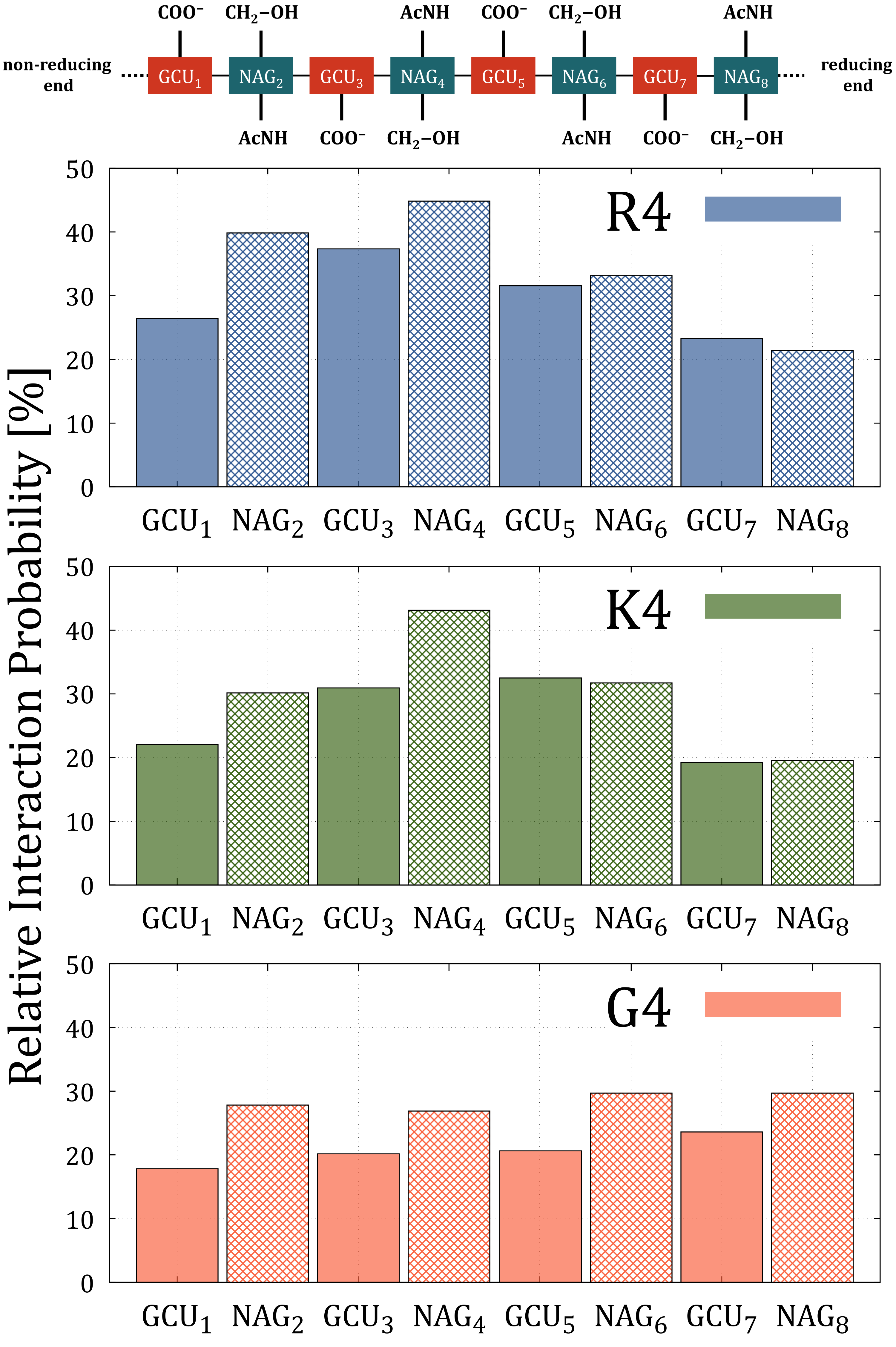}
    \caption{The relative probability of tetrapeptide binding to each of monosaccharides of the hyaluronan octamer as calculated from molecular dynamics simulations with the prosECCo75 model. The relative probability is calculated as the fraction of simulations frames with at least one hydrogen bond between tetrapeptide and monosaccharide with respect to the overall probability of tetrapeptide--hyaluronan binding as given in Figure~\ref{fgr:NMR_MD_summary}d. The probabilities for D-glucuronic acid (GCU monosaccharides) are shown as filled boxes, while the probabilities for N-acetyl-D-glucosamine (NAG monosaccharides) are shown as boxes filled with a diagonal crosshatch pattern. The schematic representation of the hyaluronan octamer highlighting its key functional groups, i.e., carboxyl COO\textsuperscript{-}, acetamide AcNH, and hydroxymethyl CH\textsubscript{2}--OH groups, is given above the graph.}
    \label{fgr:MD_bind_per_sacch}
\end{figure}

We observe a tendency of charged tetrapeptides to interact with the middle of the hyaluronan chain rather than polymer end-monosaccharides, see Figure~\ref{fgr:MD_bind_per_sacch}, which reports the relative probability of a tetrapeptide to interact with each monosaccharide of the hyaluronan octamer. This trend can indicate a possible wrapping of a charged tetrapeptide by the neighboring functional groups of hyaluronan. In the case of G4, such interaction preference is absent. Interestingly, we generally see a higher interaction probability for amino monosaccharides. This tendency can be explained by the fact that each charge--charge peptide--sugar binding event via a glucuronic acid (most often belonging to GCU3 and GCU5 monosaccharides) is often stabilized by interactions with two neighboring N-acetyl-D-glucosamines, see again Figure~\ref{fgr:NMR_shifts_MD_data}b. Another reason is that non-covalent intermolecular interactions are simply more probable for larger monosaccharides, i.e., N-acetyl-D-glucosamines, as seen in the example of G4.

\begin{figure}[htb!]
    \centering
    \includegraphics[width=\textwidth]{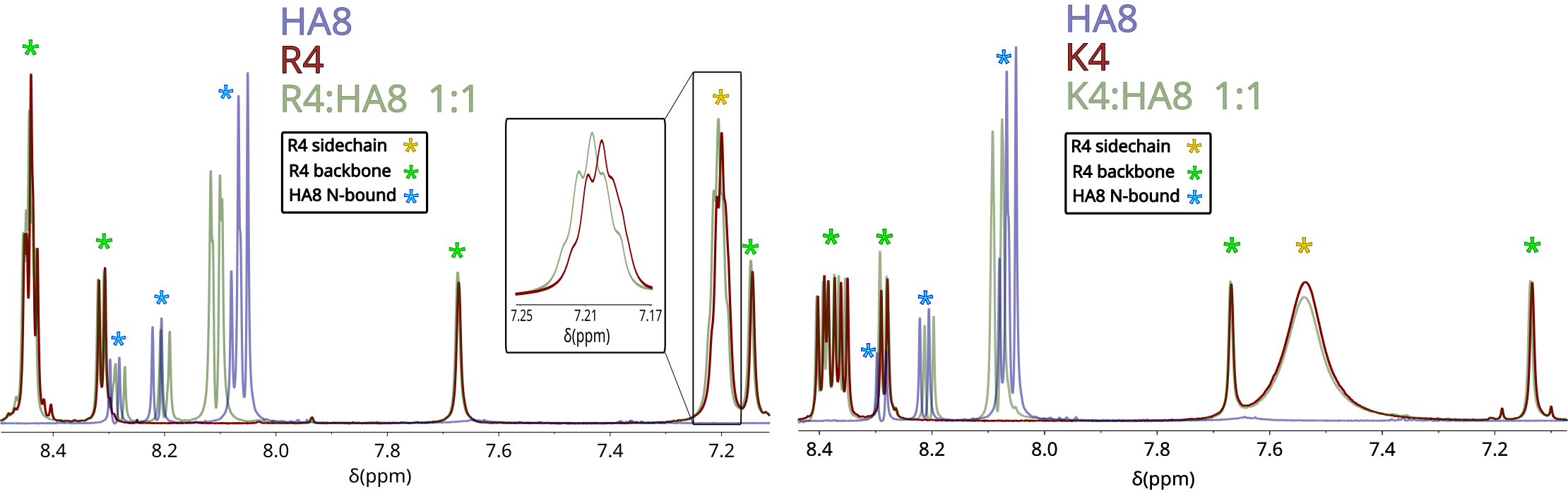}
    \caption{The region of the water-suppressed \textsuperscript{1}H NMR spectra representing N-bound protons of R4 (top) and K4 (bottom), as well as HA8 N-bound protons. The blue lines correspond to the pure HA8 solution, the red lines to pure tetrapeptide solutions, and the green lines represent the tetrapeptide--HA8 1:1 mixtures. The part of the signal with the largest perturbation, which corresponds to R4 side-chain N-bound protons, is magnified and placed into a black box.}
    \label{fgr:NMR_wat_supp}
\end{figure}

We performed water suppression NMR experiments to understand the hyaluronan interactions with tetrapeptides further. The water suppression experiments allow us to characterize changes in labile N-protons since in \textsuperscript{1}H NMR we cannot see the signal from acidic protons due to rapid hydrogen-deuterium exchange. The measurements were carried out for mixtures (i.e., R4--HA8 and K4--HA8 1:1) and pure solutions (i.e., HA8, K4, and R4). Evident modifications in the position and even the shape of HA8 N-bound signals can be observed upon adding both K4 and R4 on Figure~\ref{fgr:NMR_wat_supp}. The displacement of HA8 N-bound signal at 8.06~ppm is almost twice as large after the addition of R4 (4.8$\times10^{-2}$~ppm) than after adding K4 (2.5$\times10^{-2}$~ppm). This peak corresponds to the nitrogen of the acetamide group, which again points to its principal role in hyaluronan--tetrapeptide binding, as we previously suggested by analyzing the D\textsubscript{2}O spectra, c.f. Figure \ref{fgr:NMR_shifts_MD_data}.

At the same time, we observe almost insignificant shifts for the peptides' signal peaks in the water-suppressed \textsuperscript{1}H spectra. Somewhat notable perturbations are observed only for the R4 side-chain N-bound protons (signal at 7.20~ppm, on the order of 5$\times10^{-3}$~ppm). The signals associated with backbone N-bound protons of both tetrapeptides exhibit negligible changes. Despite supporting previous conclusions about the dominant role of basic side chains, including the higher affinity of arginine, these results indicate that the hyaluronan--peptide interactions do not alter the backbone chain structure of tetrapeptides, or at least this is not detected by our measurements.

\begin{figure}[htb!]
    \centering
    \includegraphics[width=0.7\textwidth]{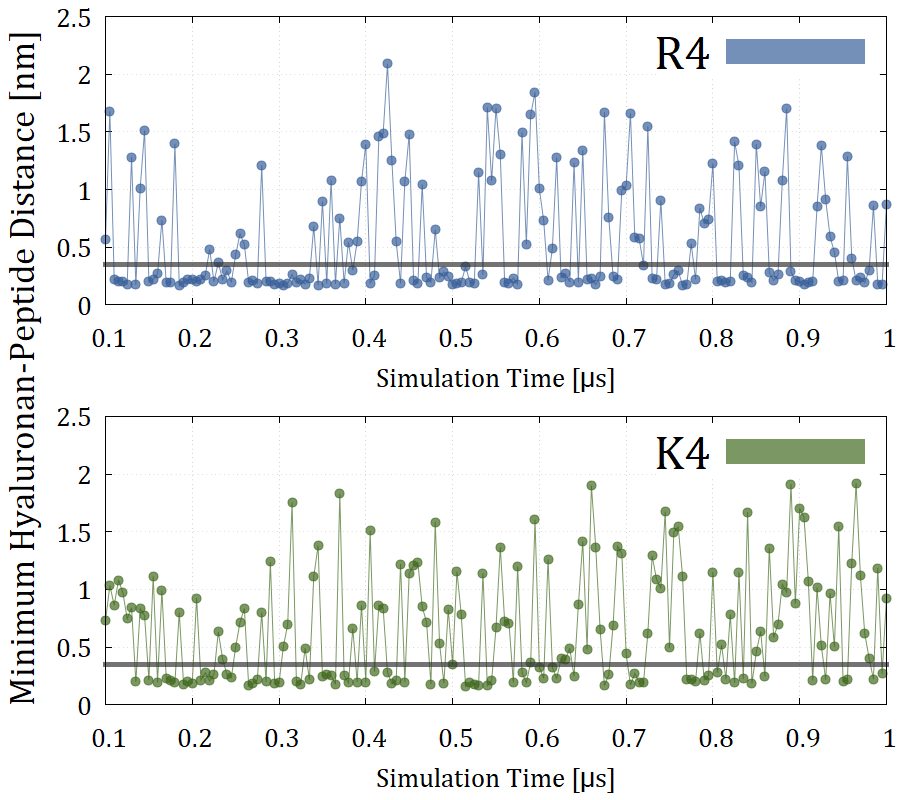}
    \caption{Time dependence of the minimum distance between hyaluronan octamer and tetrapeptides. The data are shown for HA8--R4 and HA8--K4 solutions from molecular dynamics simulations with the prosECCo75 model. The frequency of data points is 5~ns.}
    \label{fgr:MD_lifetime}
\end{figure}

In contrast to \textsuperscript{1}H NMR, no signals pointing to peptide--HA8 contacts are recorded by the water-suppressed NOESY spectra, Figures~\ref{fgr:NMR_noe_zoom_r4}, \ref{fgr:NMR_noe_zoom_k4}, \ref{fgr:NMR_NOESY_r4}, and \ref{fgr:NMR_NOESY_k4}. This observation can be explained by a low concentration of bound peptide--HA8 complexes (particularly given the low initial concentration of both solutes in our mixtures) or/and the dynamic mode of tetrapeptide--HA8 interactions, i.e., bound complexes are rather short-lived. Molecular dynamics simulations with the prosECCo75 model support the latter hypothesis. Despite relatively high interaction probabilities for R4 and K4, their binding to hyaluronan is intermittent, without forming shape-defined complexes, and with the lifetime not exceeding dozens of nanoseconds, Figure~\ref{fgr:MD_lifetime}. At the same time, the CHARMM model suggests irreversible HA8--R4 binding on the timescale of our simulations, Figure~\ref{fgr:MD_lifetime_CHARMM}, which does not align with the interpretation of our NMR data.

\section{Conclusions}

Hyaluronan is a high-molecular-weight polysaccharide vastly present in the proximal membrane region of the extracellular space. Consequently, hyaluronan--protein interactions are essential to the complex molecular picture above cellular membranes. Understanding and controlling these interactions is necessary for advancing biological and medical applications, including drug development and disease diagnosis.

This work elaborates on possible molecular mechanisms of hyaluronan--protein binding by examining hyaluronan--peptide interactions using NMR measurements and molecular dynamics simulations. Our results indicate the higher affinity of positively charged tetramers for hyaluronan, as could be expected due to its overall negative charge. Both methodologies agree on the defining role of electrostatics in these interactions and the specific side-chain affinity of basic amino acids. The chemical shift perturbation measurements show that adding R4 and K4 alters the \textsuperscript{1}H NMR spectra of hyaluronan while adding charge-neutral G4 does not result in any remarkable changes. R4-induced signal displacements are larger than those induced by K4, while both tetrapeptides affect the same hyaluronan protons, suggesting a similar interaction mechanism. This mechanism, supported by our MD simulations, proposes non-bonded interactions involving charged groups of hyaluronan and tetrapeptides, which are also stabilized by non-covalent binding via the acetamide and hydroxymethyl groups of hyaluronan.

The water suppression experiments also detect considerable perturbations on HA8 N-bound protons, with larger displacement for R4, further promoting the importance of the acetamide group in peptide--HA binding. The active role of the acetamide, together with its proximity to the carboxyl group, suggests a possible cooperative effect by clamping the peptide close to the polysaccharide. At the same time, only minor modifications can be observed for the peptide signals, with notable displacements only for arginine side-chain N-bound protons. These results confirm the leading role of the arginine side chain in hyaluronan--peptide interactions. Additionally, the frequency of binding events seen in MD simulations and the lack of water-suppressed NOESY signals suggest that the peptide--HA interaction mode is mainly dynamic, with regular but short-lived contacts, without forming shape-defined, irreversibly bound hyaluronan--tetrapeptides complexes.

Our data systematize previously reported results, where arginine residues were shown to be essential for hyaluronan interactions with proteins, including its main transmembrane receptor CD44 \cite{Vuorio2017}. Considering occasionally contradicting reports about the strength and nature of hyaluronan interactions, our data can help resolve these controversies by providing a plausible binding molecular mechanism that relies on electrostatics and amino acid selectivity. At the biological timescale, hyaluronan's intermittent but frequent interactions with proteins are credible. Otherwise, hyaluronan polymers, extremely large in length and hydrodynamic volume, would affect the protein transport toward and along cellular membranes. At the same time, hyaluronan may play a crucial role in stabilizing and guiding protein or ligand binding to lipid membranes, glycosaminoglycans, and other macromolecules in the extracellular matrix.

\begin{acknowledgement}

M.R. and H.M.-S. acknowledge the support of the Czech Science Foundation (project 19-19561S). M.R. also acknowledges the support from the Charles University in Prague and the International Max Planck Research School in Dresden. We acknowledge Grammarly and ChatGPT 3.5 for improving readability and language. 

\end{acknowledgement}

\bibliography{library.bib}

\clearpage
\setcounter{figure}{0}
\begin{suppinfo}

\renewcommand{\thefigure}{S\arabic{figure}}
\renewcommand{\thetable}{S\arabic{table}}

\subsubsection{NMR reports}

\textbf{HA8 \textsuperscript{1}H NMR (401 MHz, D\textsubscript{2}O):} $\delta$ 4.61 – 4.49 (m, 1H, (NAG1)), 4.46 (d, J = 7.8 Hz, 1H, (GCU1)), 3.96 – 3.85 (m, 1H, (NAG6a)), 3.85 – 3.81 (m, 1H, (NAG2)), 3.81 – 3.67 (m, 4H,(NAG3,NAG6b,GCU4,GCU5)), 3.64 – 3.43 (m, 3H, (NAG5,NAG4,GCU3)), 3.34 (m, J = 8.7 Hz, 1H, (GCU2)), 2.08 – 1.97 (m, 3H, (NAG Ac)). 401 MHz, D2O).

{\setlength{\parindent}{0pt}
\textbf{R4 \textsuperscript{1}H NMR (401 MHz, D\textsubscript{2}O):} $\delta$ 4.40 – 4.18 (m, 4H, CH(a)), 3.21 (m, J = 6.8, 3.7 Hz, 8H, CH\textsubscript{2}(d)), 2.03 (s, J = 0.7 Hz, 3H, Ac), 1.94 – 1.55 (m, 16H, CH\textsubscript{2}(b,c)).

\textbf{K4 \textsuperscript{1}H NMR (401 MHz, D\textsubscript{2}O):} $\delta$ 4.42 – 4.20 (m, 4H, CH(a)), 3.00 (t, J = 7.7 Hz, 8H, CH\textsubscript{2}(e)), 2.03 (s, J = 0.9 Hz, 3H, Ac), 1.94 – 1.61 (m, 16H, CH\textsubscript{2}(b,d)), 1.44 (m, J = 15.3, 7.7 Hz, 8H, CH\textsubscript{2}(c)).

%working on this one
\textbf{G4 \textsuperscript{1}H NMR (401 MHz, D\textsubscript{2}O):} $\delta$ 4.00 (d, J = 3.2 Hz, 4H, CH(b)), 3.97 (s, 2H, CH(a)), 3.93 (s, 2H, CH(c)), 2.07 (s, 3H, Ac).}

\clearpage

\subsubsection{Additional NMR spectra}

\begin{figure}[htb!]
    \centering
    \includegraphics[width=\textwidth]{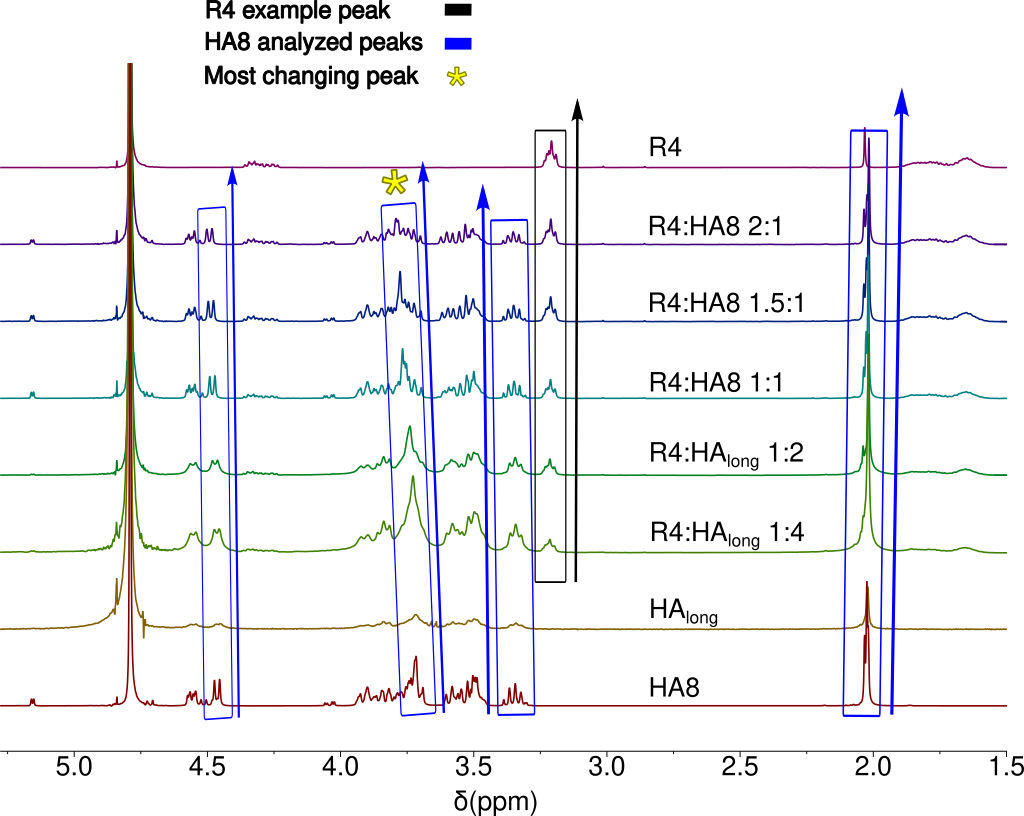}
    \caption{\textsuperscript{1}H NMR spectra for the collection of R4--hyaluronan solutions. The blue boxes and blue arrows guide through the displacement of the analyzed HA peaks upon the addition of R4, while the black box and black arrow point to the lack of changes in the peaks originating from the peptide.}
    \label{fgr:NMR_R4_H8_1H}
\end{figure}
\clearpage

\begin{figure}[htb!]
    \centering
    \includegraphics[width=\textwidth]{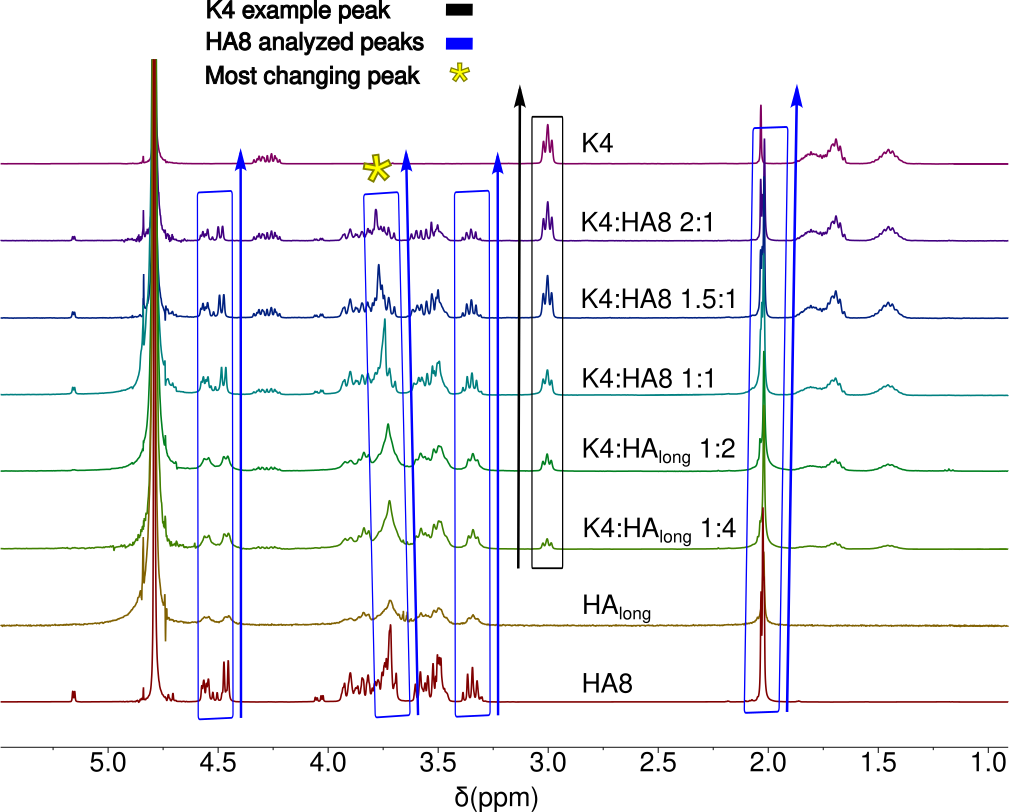}
    \caption{\textsuperscript{1}H NMR spectra for the collection of K4--hyaluronan solutions. The blue boxes and blue arrows guide through the displacement of the analyzed HA peaks upon the addition of K4, while the black box and black arrow point to the lack of changes in the peaks originating from the peptide.}
    \label{fgr:NMR_K4_H8_1H}
\end{figure}
\clearpage

\begin{figure}[htb!]
    \centering
    \includegraphics[width=\textwidth]{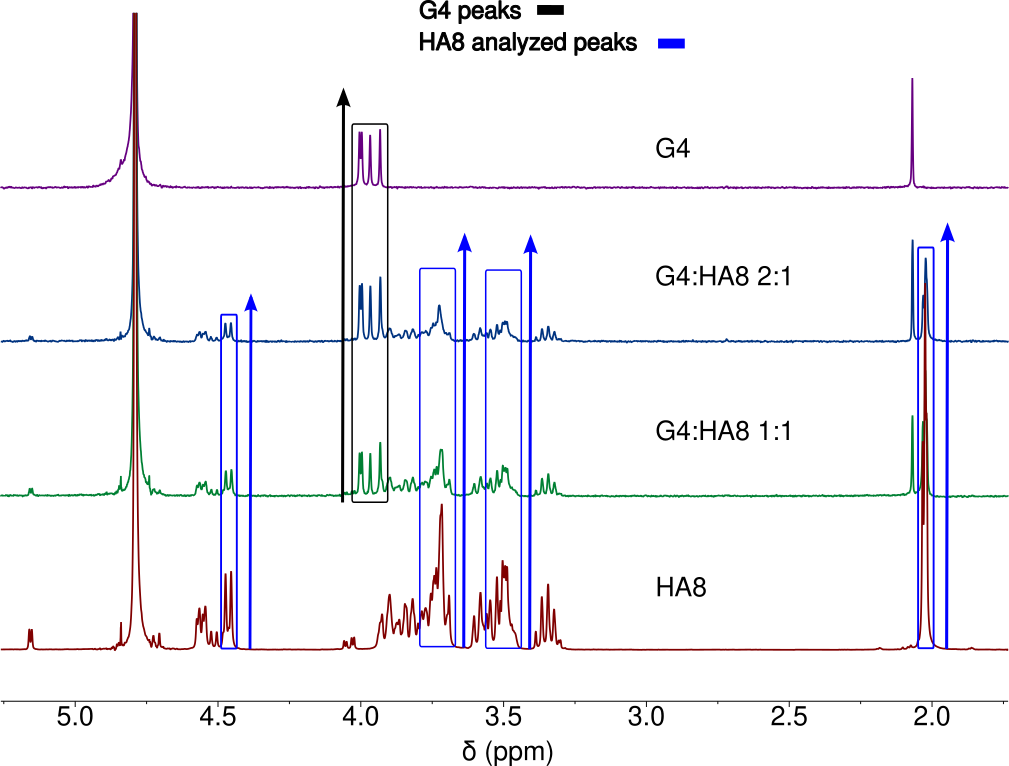}
    \caption{\textsuperscript{1}H NMR spectra for the collection of G4--hyaluronan solutions. The blue boxes and blue arrows guide through the lack of changes in the analyzed HA peaks upon the addition of G4, while the black box and black arrow point to the lack of changes in the peaks originating from the peptide.}
    \label{fgr:NMR_G4_H8_1H}
\end{figure}
\clearpage

\begin{figure}[htb!]
    \centering
    \includegraphics[width=\textwidth]{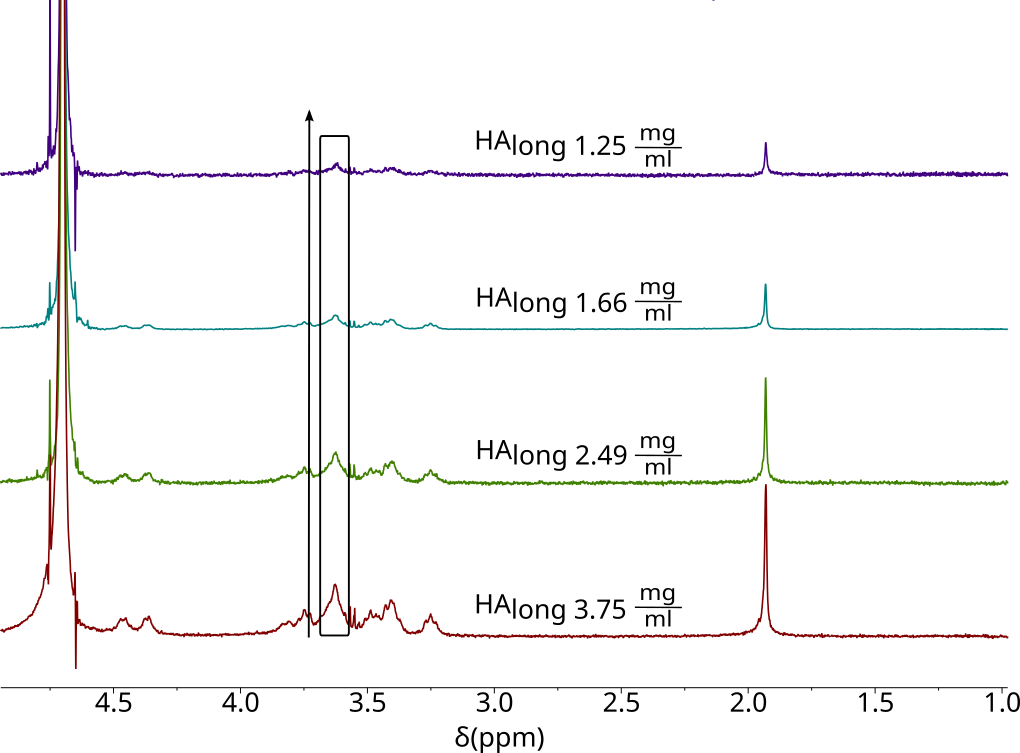}
    \caption{The comparison of the \textsuperscript{1}H NMR spectra at different concentrations of the large hyaluronan (HA\textsubscript{long}) without the addition of any peptide. The position of the signal with the largest changes upon R4/K4 addition in the other experiments is placed in a black box. The slope of its concentration-dependent change is represented with an arrow, i.e., we observe no displacement of the chemical shift. The concentrations mentioned in the figure are equivalent to those used in the peptide titration experiments with HA8. 3.75~mg/ml corresponds to the concentration of the HA8 alone, 2.49~mg/ml to the 1:1 peptide:HA8 ratio, 1.66~mg/ml to the 1:1.5 ratio, and 1.25~mg/ml to the 1:2 ratio.}
    \label{fgr:NMR_long_HA_conc}
\end{figure}
\clearpage

\begin{figure}[htb!]
    \centering
    \includegraphics[width=\textwidth]{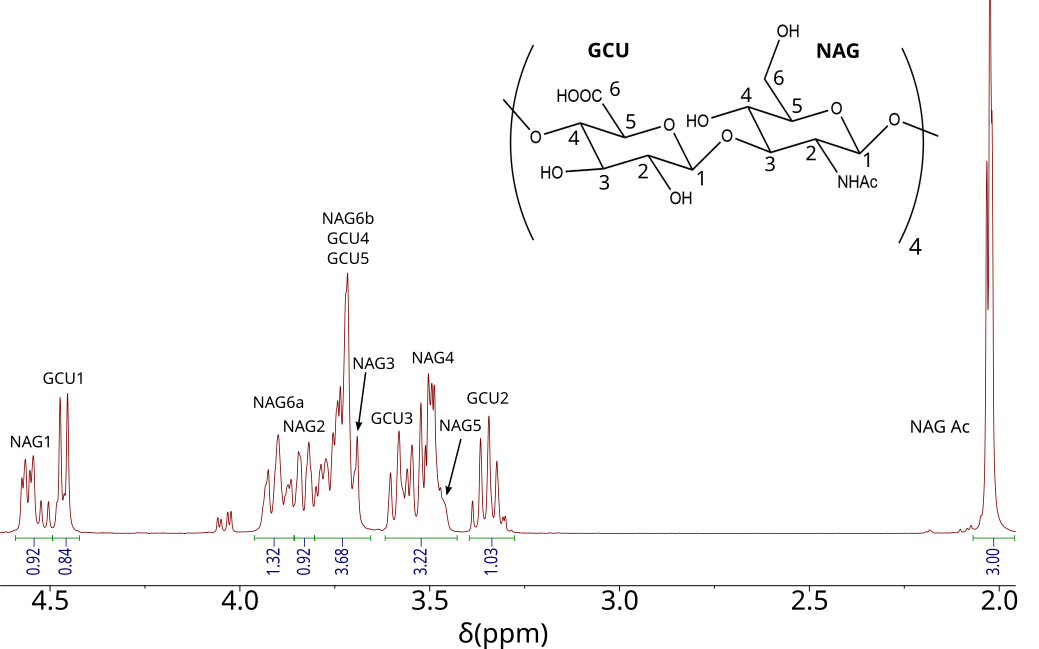}
    \caption{Signal assignment for \textsuperscript{1}H NMR spectrum of hyaluronan octamer (HA8).}
    \label{fgr:NMR_assign_ha}
\end{figure}
\clearpage

\begin{figure}[htb!]
    \centering
    \includegraphics[width=\textwidth]{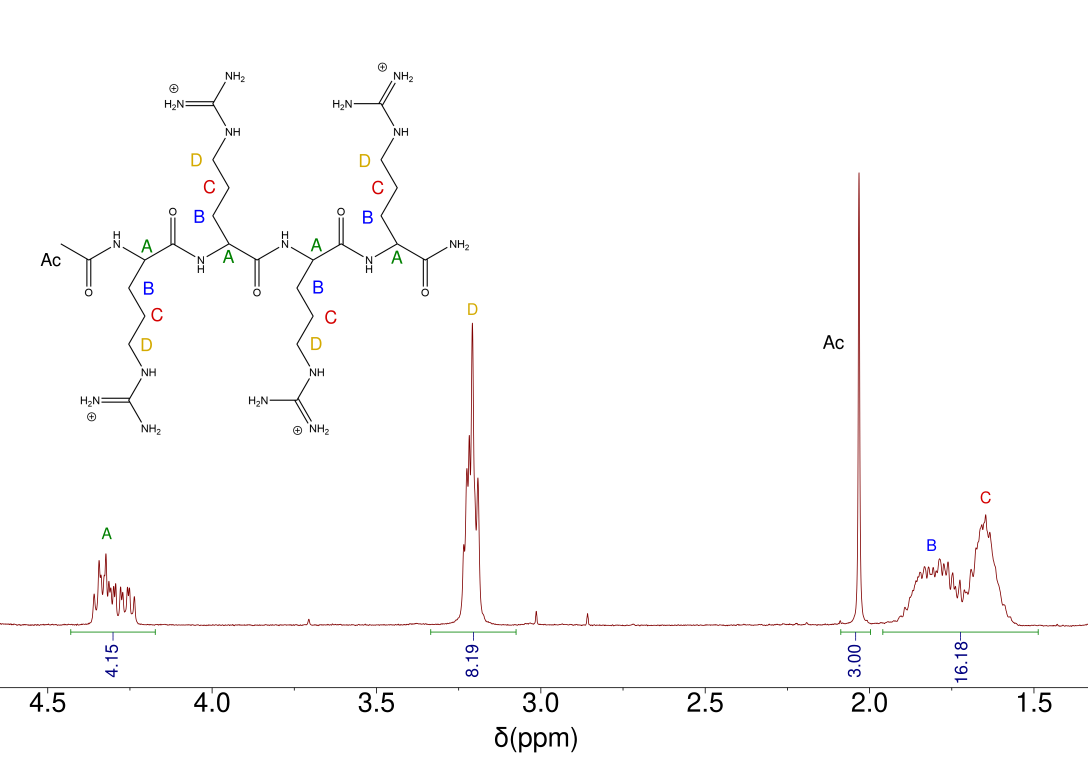}
    \caption{Signal assignment for \textsuperscript{1}H NMR spectrum of tetraarginine (R4).}
    \label{fgr:NMR_assign_r4}
\end{figure}
\clearpage

\begin{figure}[htb!]
    \centering
    \includegraphics[width=\textwidth]{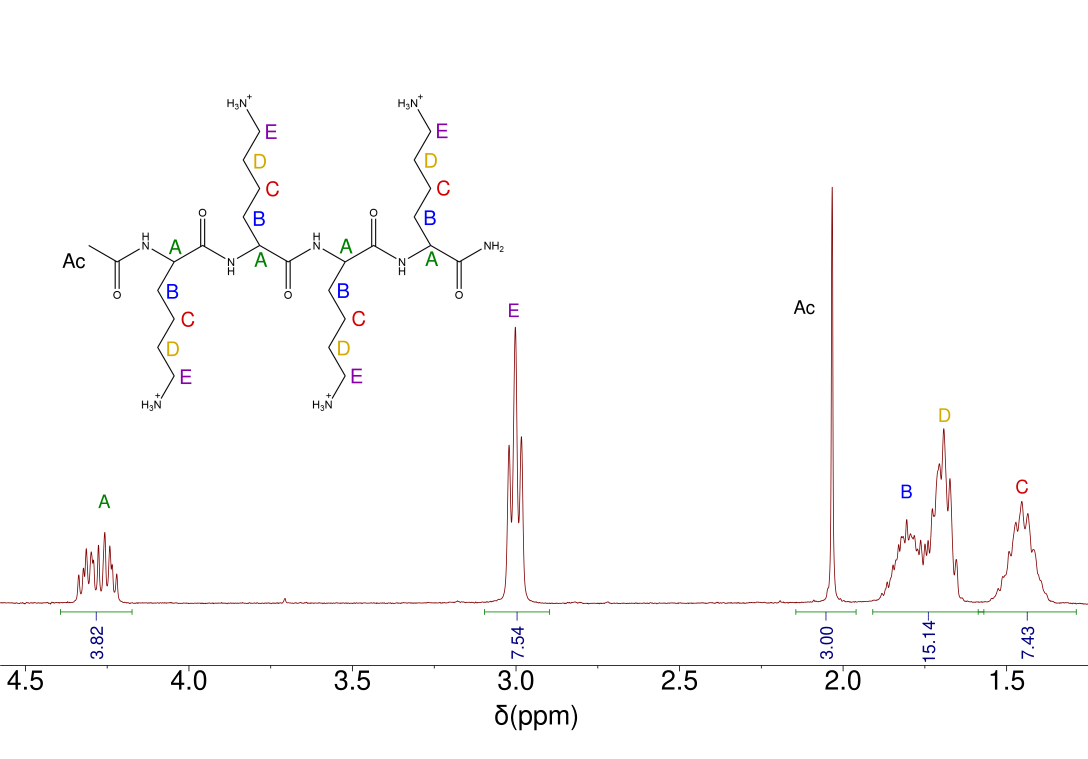}
    \caption{Signal assignment for \textsuperscript{1}H NMR spectrum of tetralysine (K4).}
    \label{fgr:NMR_assign_k4}
\end{figure}
\clearpage

\begin{figure}[htb!]
    \centering
    \includegraphics[width=\textwidth]{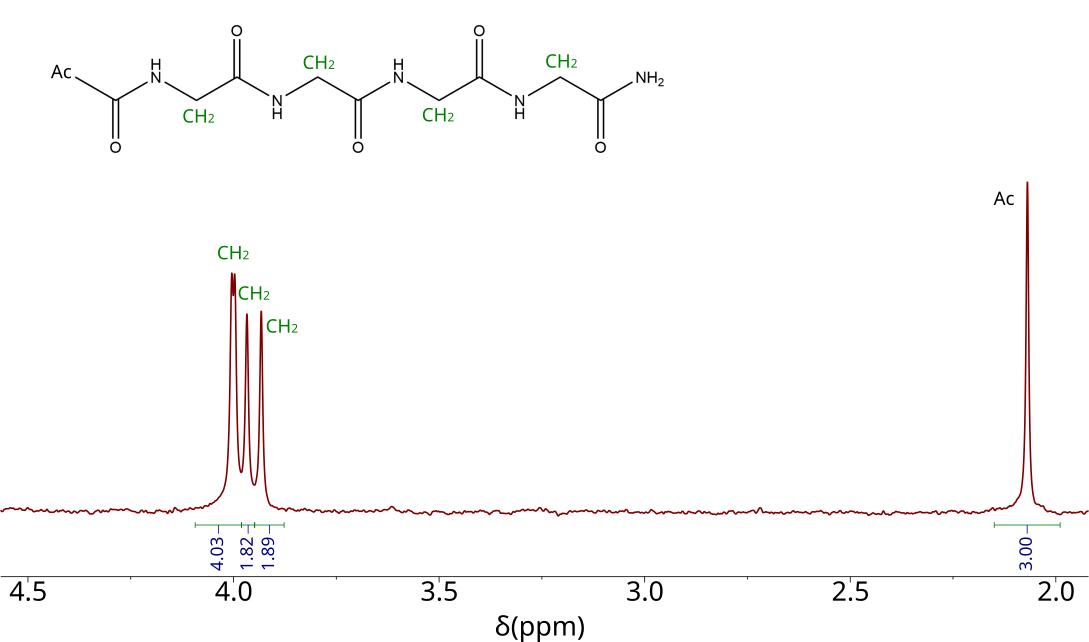}
    \caption{Signal assignment for \textsuperscript{1}H NMR spectrum of tetraglycine (G4). Note that the CH\textsubscript{2} groups of G4 are indistinguishable in \textsuperscript{1}H NMR spectrum.}
    \label{fgr:NMR_assign_g4}
\end{figure}
\clearpage

\begin{figure}[htb!]
    \centering
    \includegraphics[width=\textwidth]{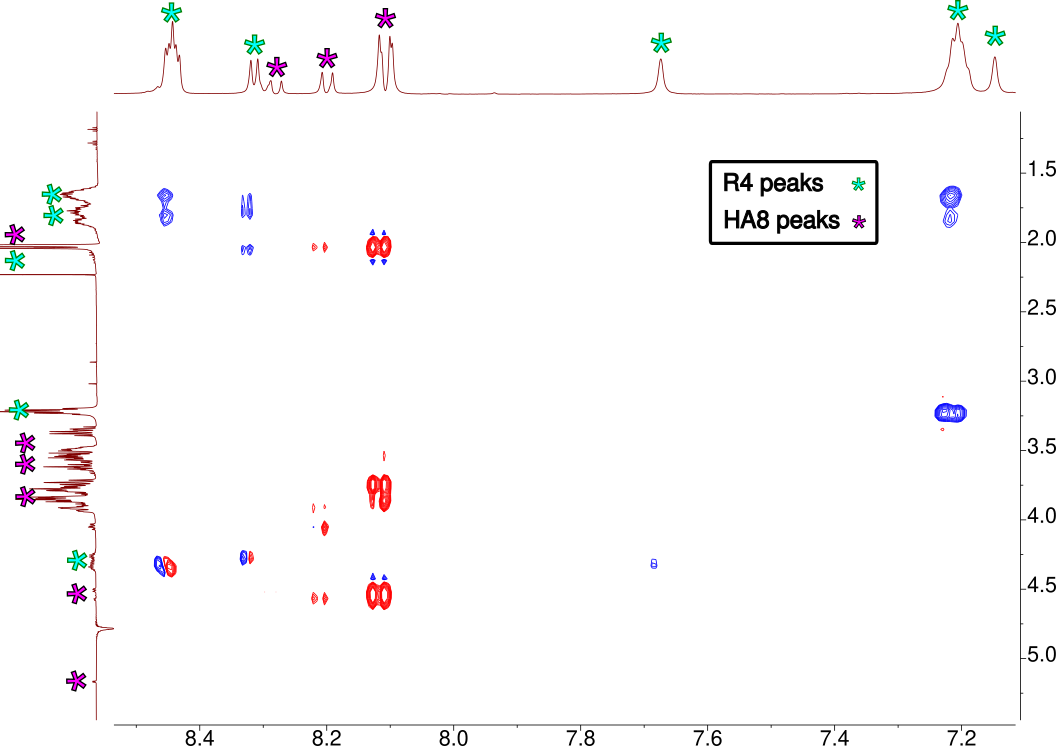}
    \caption{NOESY spectrum for the R4:HA8 solution from the water suppression experiments. Only the region with NOE signals between N-bound and non-N-bound protons is shown.}
    \label{fgr:NMR_noe_zoom_r4}
\end{figure}
\clearpage

\begin{figure}[htb!]
    \centering
    \includegraphics[width=\textwidth]{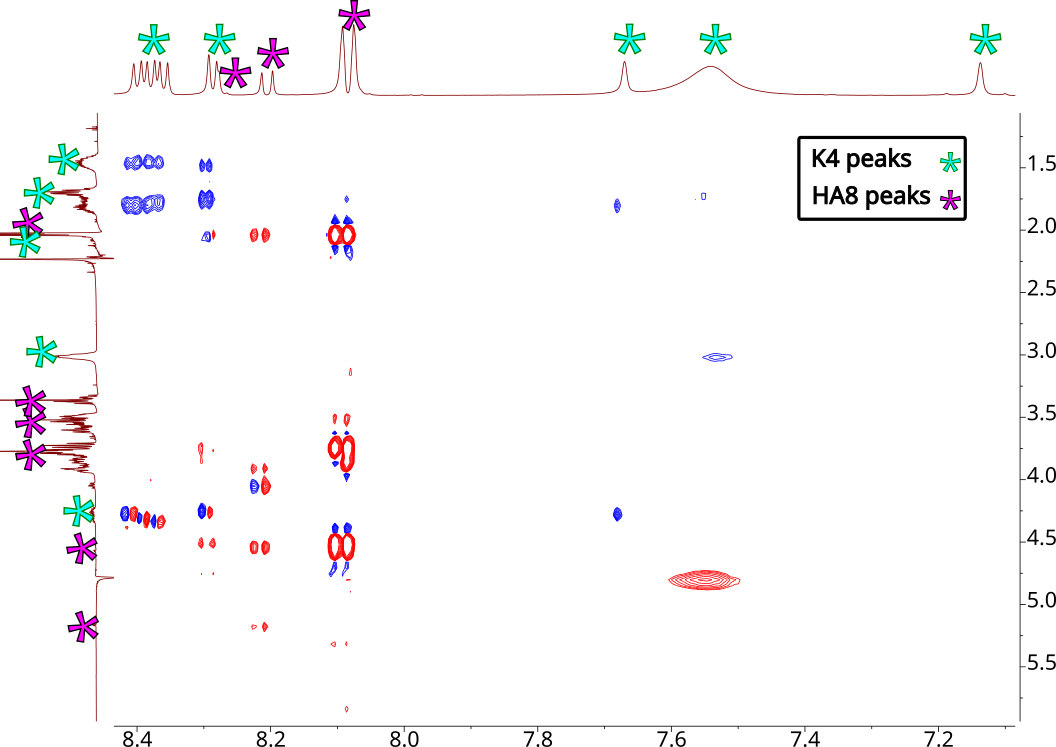}
    \caption{NOESY spectrum for the K4:HA8 solution from the water suppression experiments. Only the region with NOE signals between N-bound and non-N-bound protons is shown.}
    \label{fgr:NMR_noe_zoom_k4}
\end{figure}
\clearpage

\begin{figure}[htb!]
    \centering
    \includegraphics[width=\textwidth]{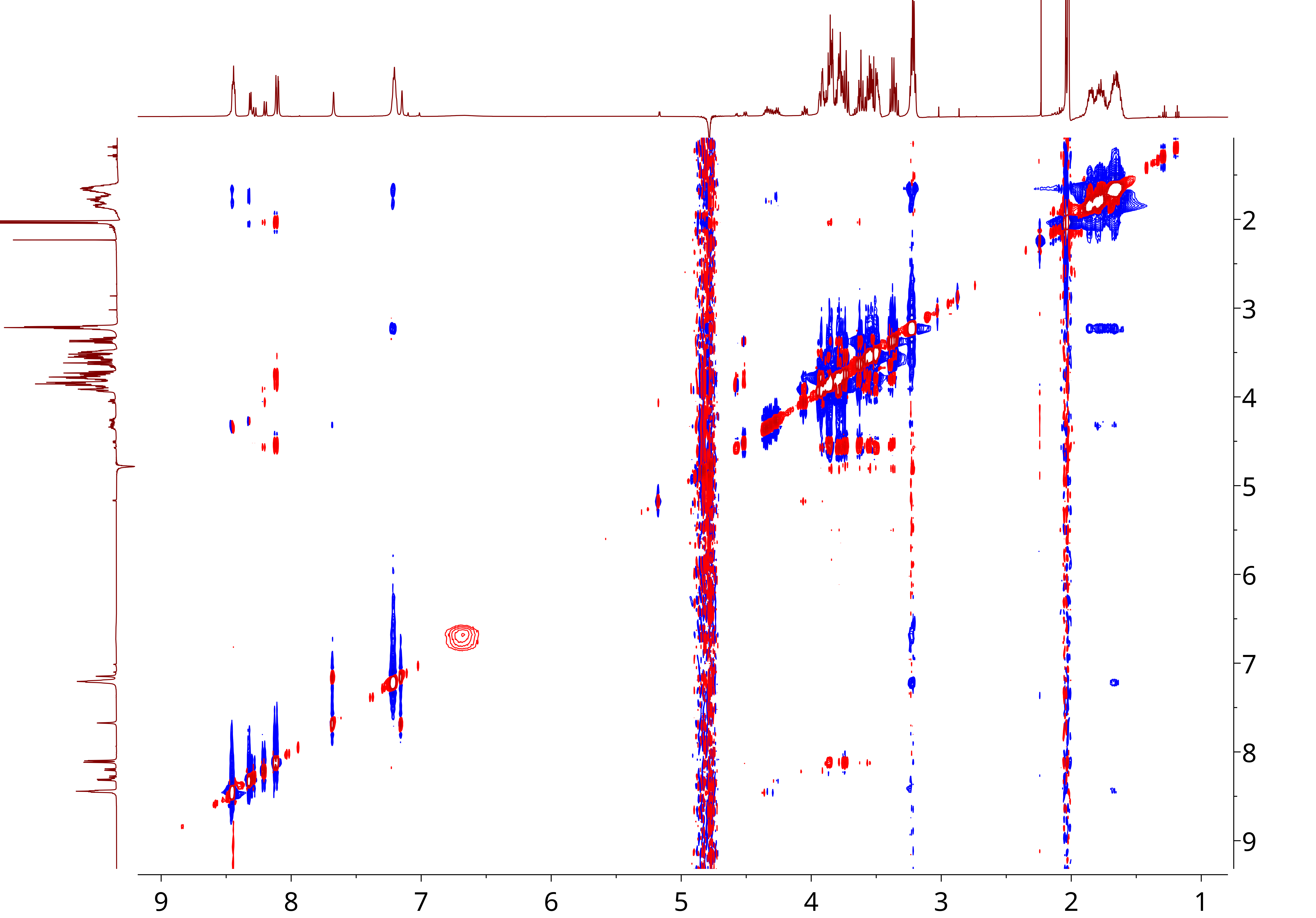}
    \caption{Full NOESY spectrum for the R4:HA8 1:1 solution measured from the water suppression experiments.}
    \label{fgr:NMR_NOESY_r4}
\end{figure}
\clearpage

\begin{figure}[htb!]
    \centering
    \includegraphics[width=\textwidth]{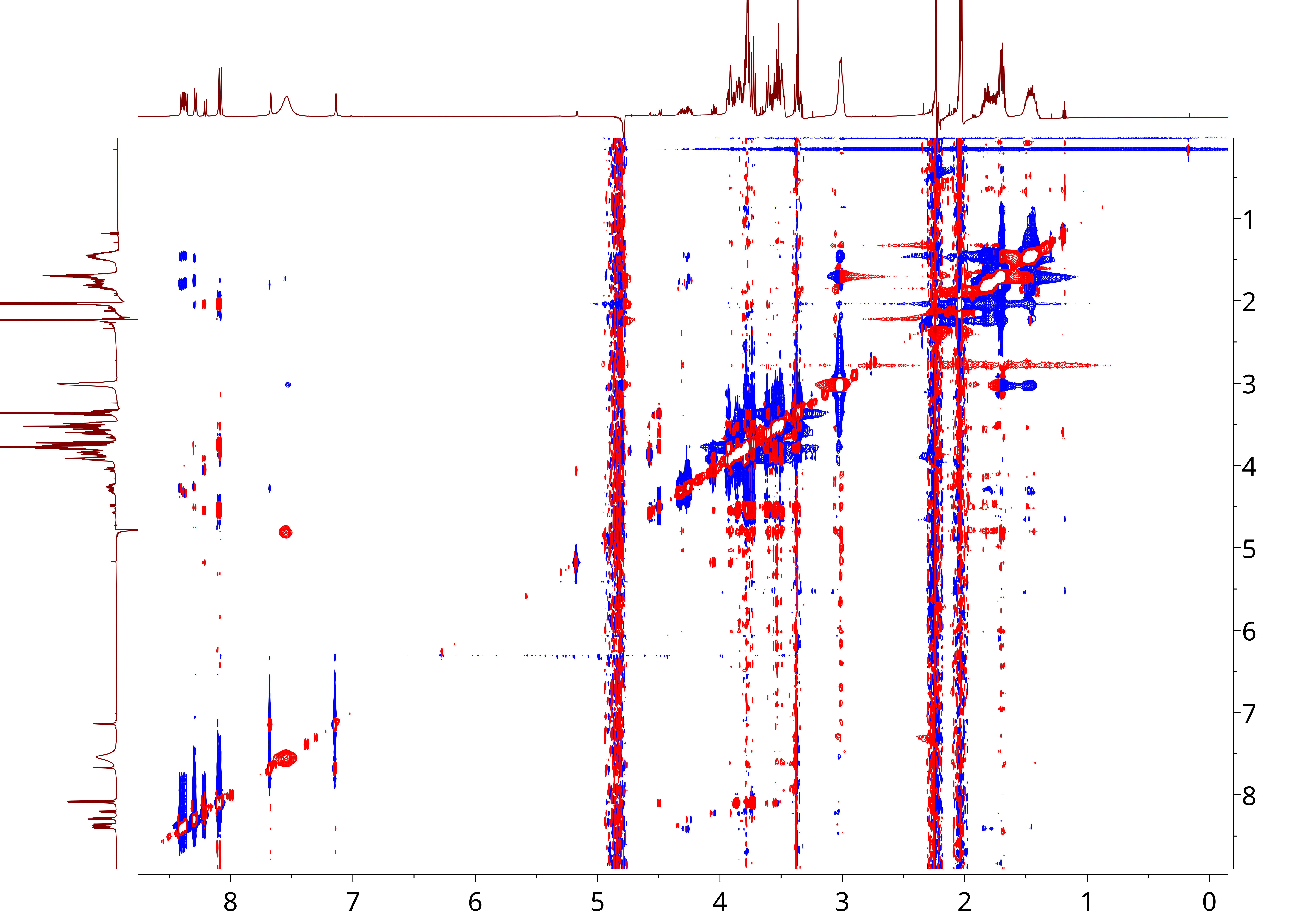}
    \caption{Full NOESY spectrum for the K4:HA8 1:1 solution measured from the water suppression experiments.}
    \label{fgr:NMR_NOESY_k4}
\end{figure}
\clearpage

\begin{figure}[htb!]
    \centering
    \includegraphics[width=\textwidth]{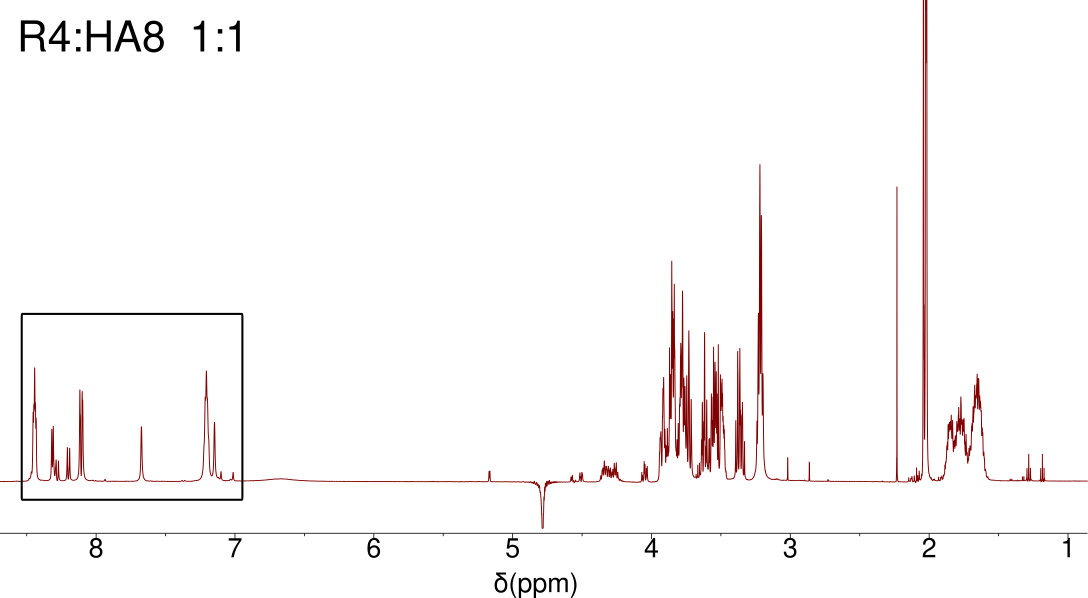}
    \caption{Full \textsuperscript{1}H spectrum for the R4:HA8 1:1 solution measured from the water suppression experiments. The part of the signal that belongs to N-labile protons and commented in detail in the main text is shown inside a black box.}
    \label{fgr:NMR_water_suppressed_full_r4}
\end{figure}
\clearpage

\begin{figure}[htb!]
    \centering
    \includegraphics[width=\textwidth]{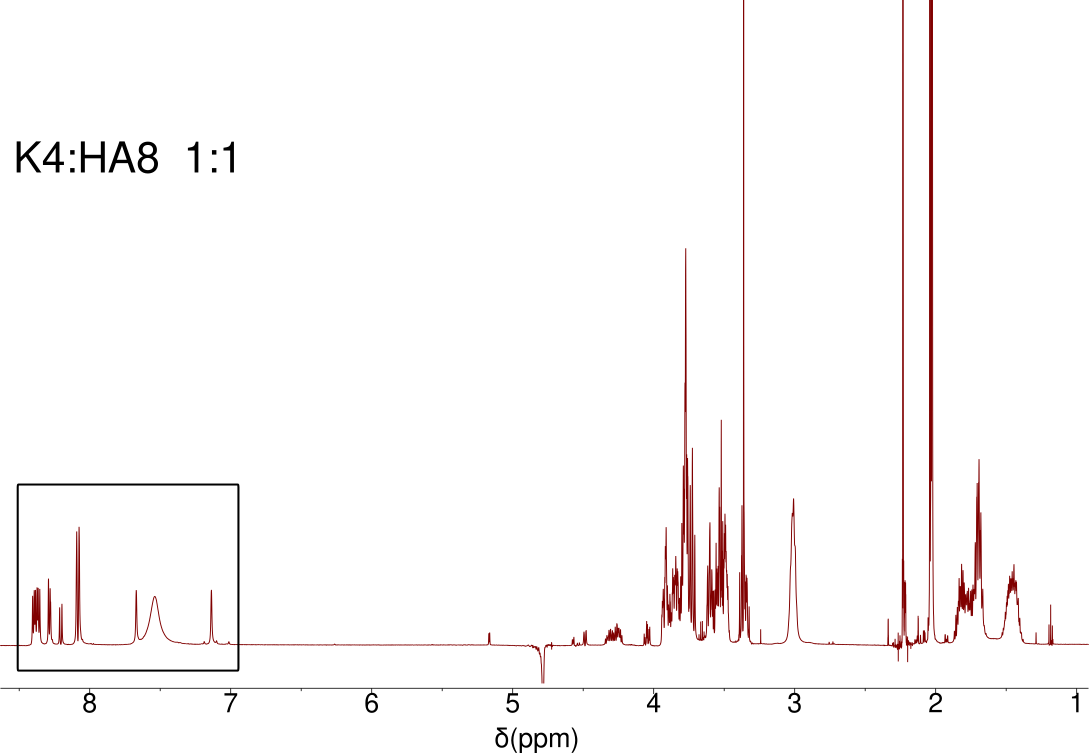}
    \caption{Full \textsuperscript{1}H spectrum for the K4:HA8 1:1 solution measured from the water suppression experiments. The part of the signal that belongs to N-labile protons and commented in detail in the main text is shown inside a black box.}
    \label{fgr:NMR_water_suppressed_full_K4}
\end{figure}
\clearpage

% Adding bidimensional spectra:

\begin{figure}[htb!]
    \centering
    \includegraphics[width=\textwidth]{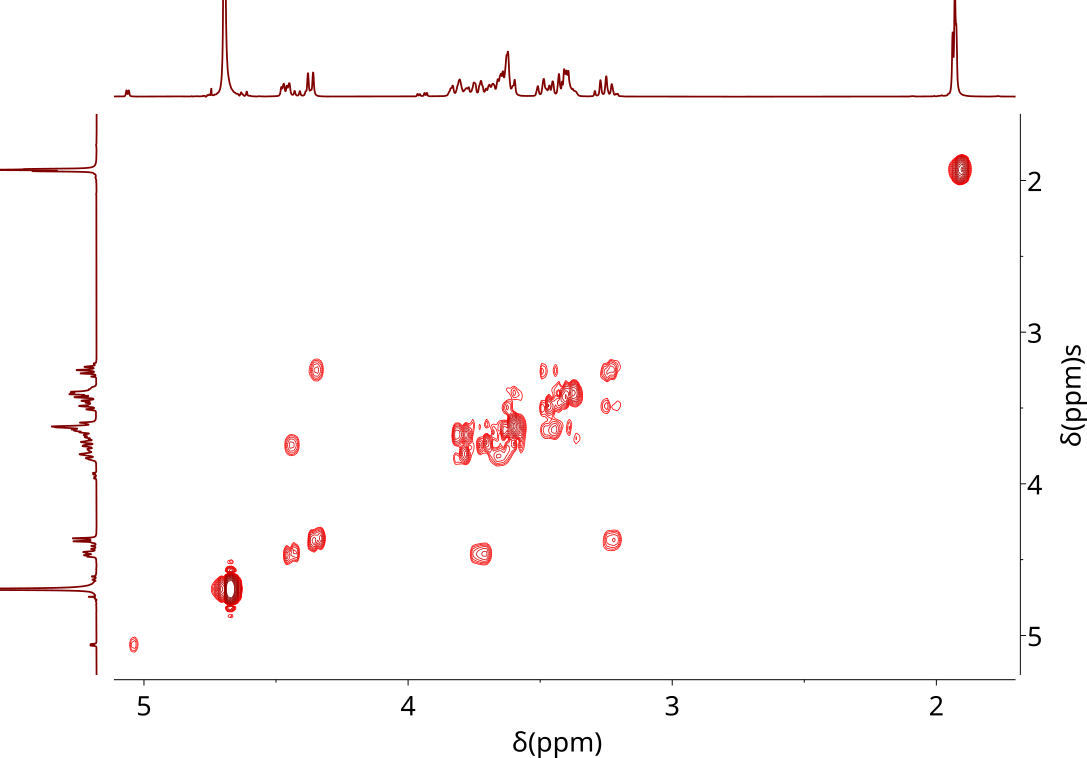}
    \caption{HA8 COSY spectrum.}
\end{figure}
\clearpage

\begin{figure}[htb!]
    \centering
    \includegraphics[width=\textwidth]{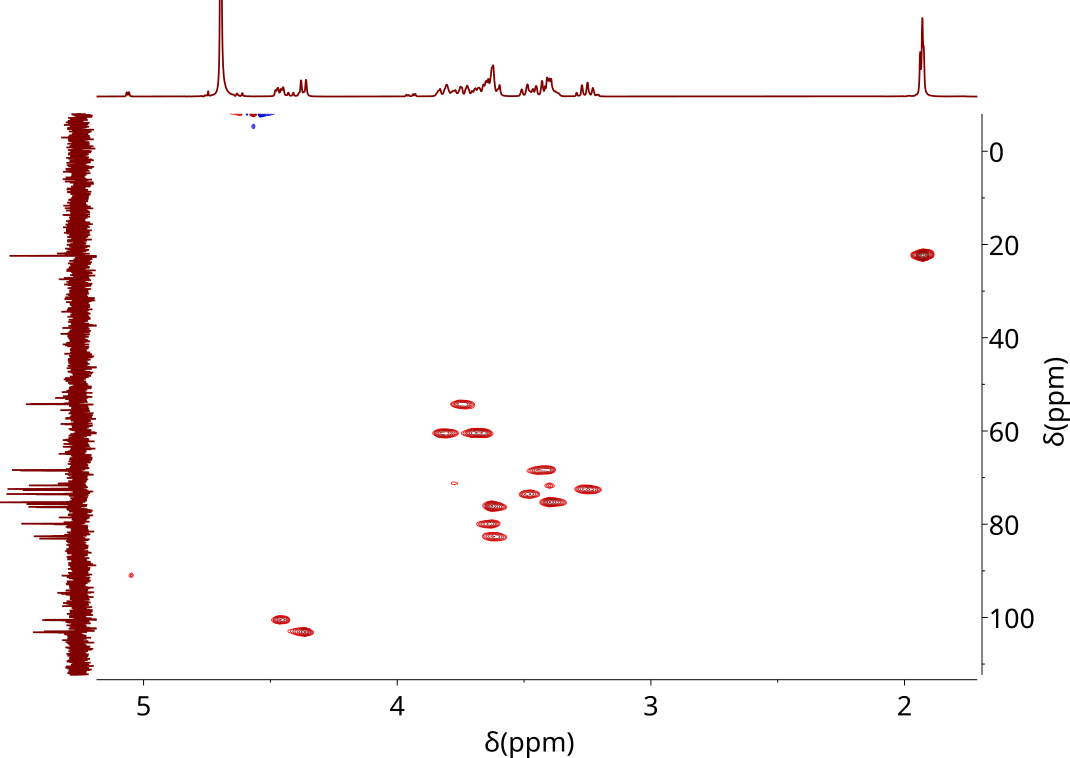}
    \caption{HA8 HSQC spectrum.}
\end{figure}
\clearpage

\begin{figure}[htb!]
    \centering
    \includegraphics[width=\textwidth]{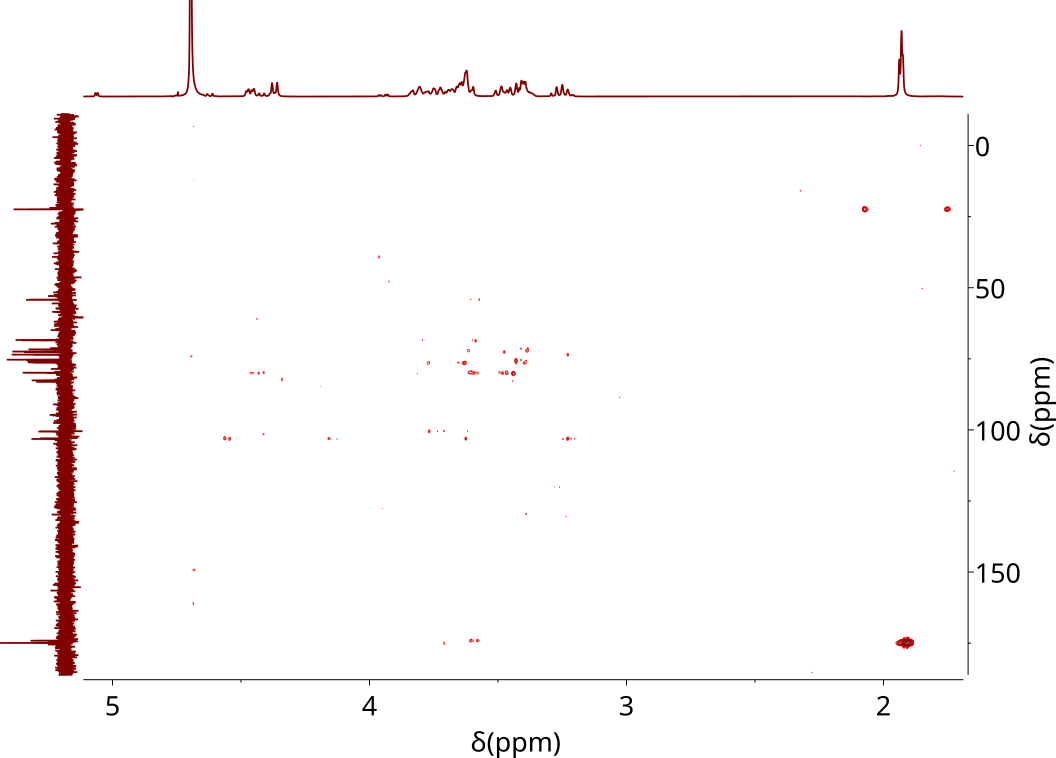}
    \caption{HA8 HMBC spectrum.}
\end{figure}
\clearpage

\begin{figure}[htb!]
    \centering
    \includegraphics[width=\textwidth]{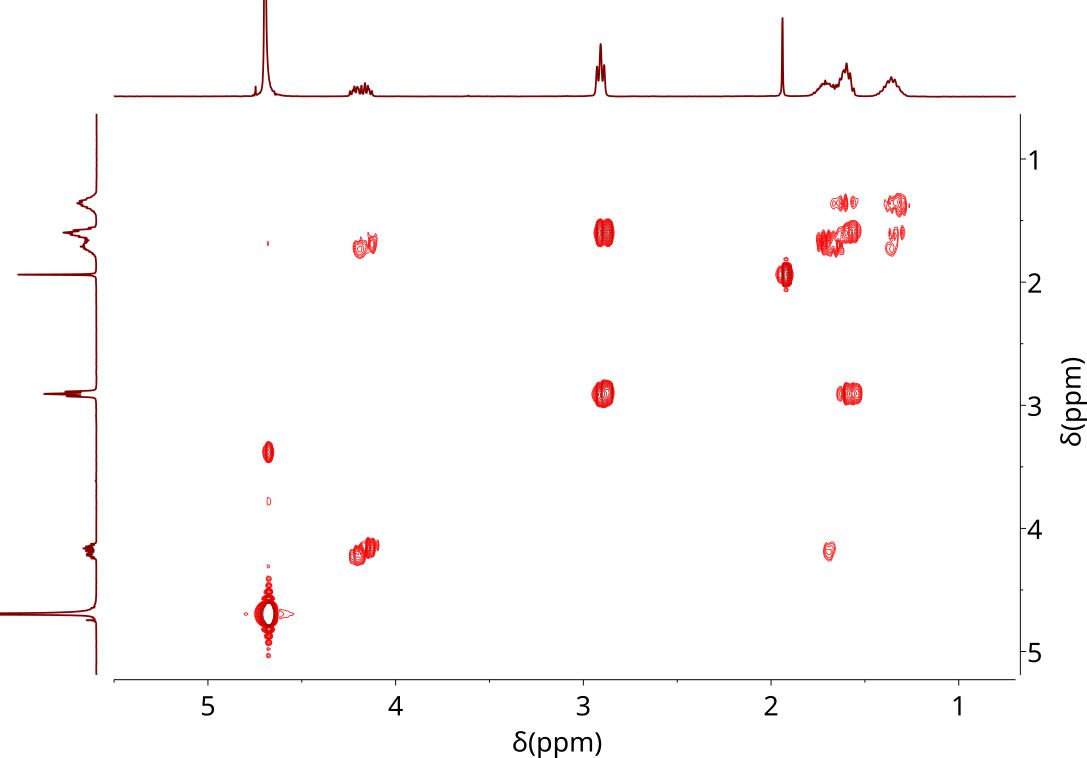}
    \caption{K4 COSY spectrum.}
\end{figure}
\clearpage

\begin{figure}[htb!]
    \centering
    \includegraphics[width=\textwidth]{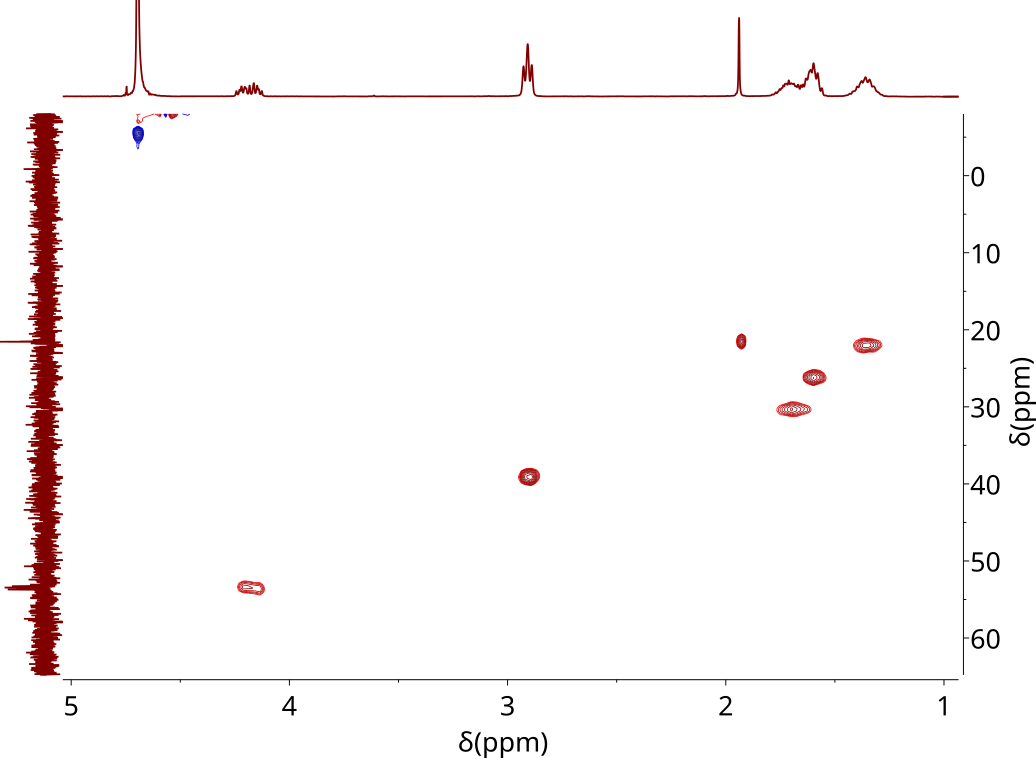}
    \caption{K4 HSQC spectrum.}
\end{figure}
\clearpage

\begin{figure}[htb!]
    \centering
    \includegraphics[width=\textwidth]{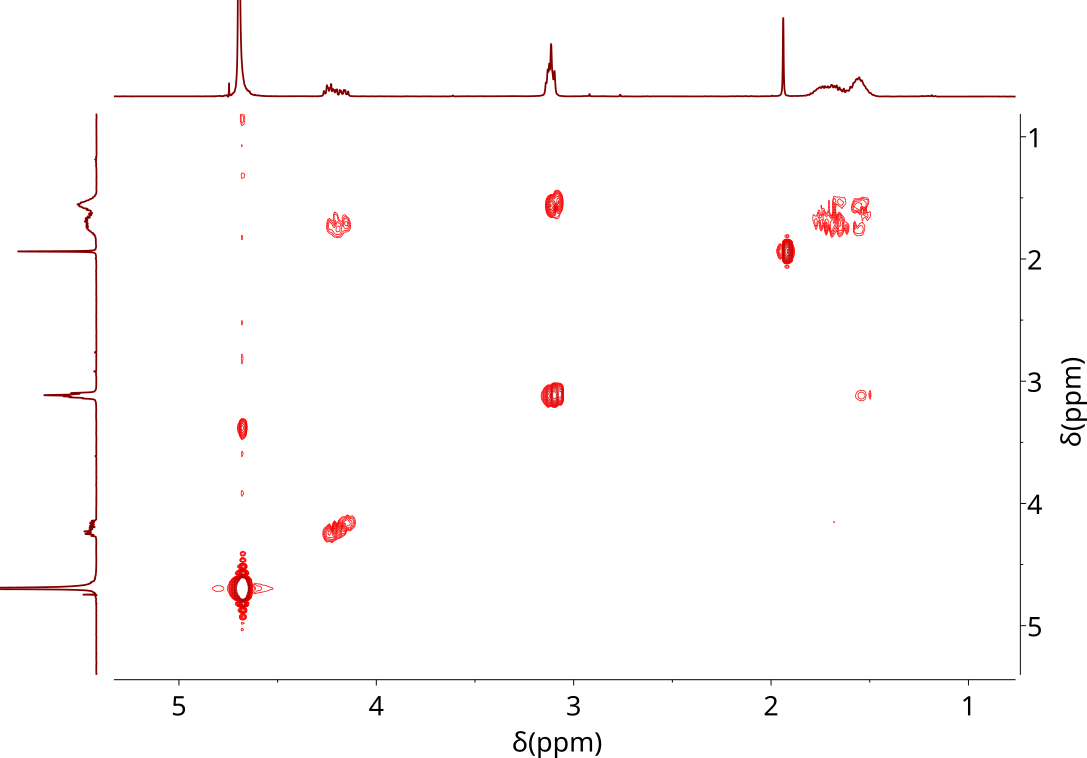}
    \caption{R4 COSY spectrum.}
\end{figure}
\clearpage

\begin{figure}[htb!]
    \centering
    \includegraphics[width=\textwidth]{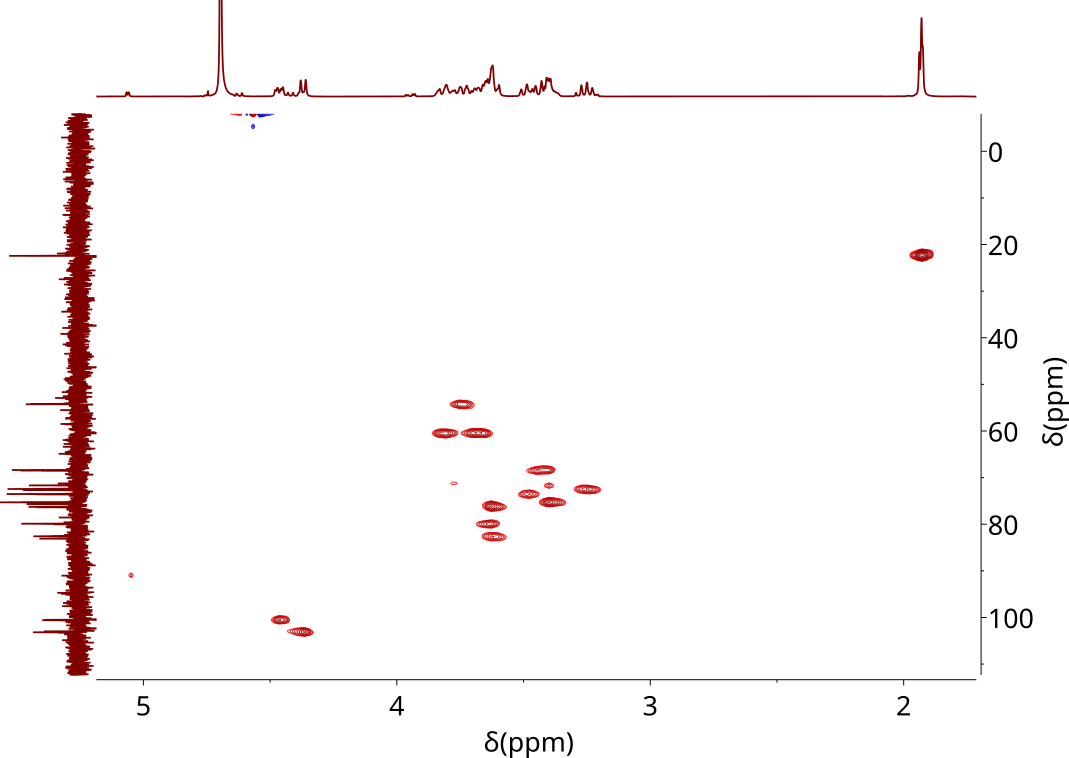}
    \caption{R4 HSQC spectrum.}
\end{figure}
\clearpage

\subsubsection{Additional Molecular Dynamics Details and Results}

\begin{table}[ht]
    \caption{The partial atomic charges (CHARMM nomenclature) of ionic groups of hyaluronan (HA8), tetraarginine (R4), and tetralysine (K4) from CHARMM and prosECCo75 force fields}
    \centering
    \begin{tabular}{ c c r r }
    Molecule & Atom Name(s) & CHARMM & prosECCo75 \\
    \hline
    \multirow{2}{*}{HA8} & C6       &  0.52 &  0.39 \\
                         & O61, O62 & -0.76 & -0.57 \\
    \hline
    \multirow{7}{*}{R4} & CD                     &  0.20 &  0.15   \\
                        & HD1, HD2               &  0.09 &  0.0675 \\
                        & NE                     & -0.7  & -0.525  \\
                        & HE                     &  0.44 &  0.33   \\
                        & CZ                     &  0.64 &  0.48   \\
                        & NH1, NH2               & -0.80 & -0.60   \\
                        & HH11, HH12, HH21, HH22 &  0.46 &  0.345  \\
    \hline
    \multirow{4}{*}{K4} & CE                     &  0.21 &  0.1575 \\
                        & HE1, HE2               &  0.05 &  0.0375 \\
                        & NZ                     & -0.3  & -0.225  \\
                        & HZ1, HZ2, HZ3          &  0.33 &  0.2475 \\
    \end{tabular}
    \label{tbl:force_field_charges}
\end{table}
\clearpage

\begin{figure}[htb!]
    \centering
    \includegraphics[width=0.65\textwidth]{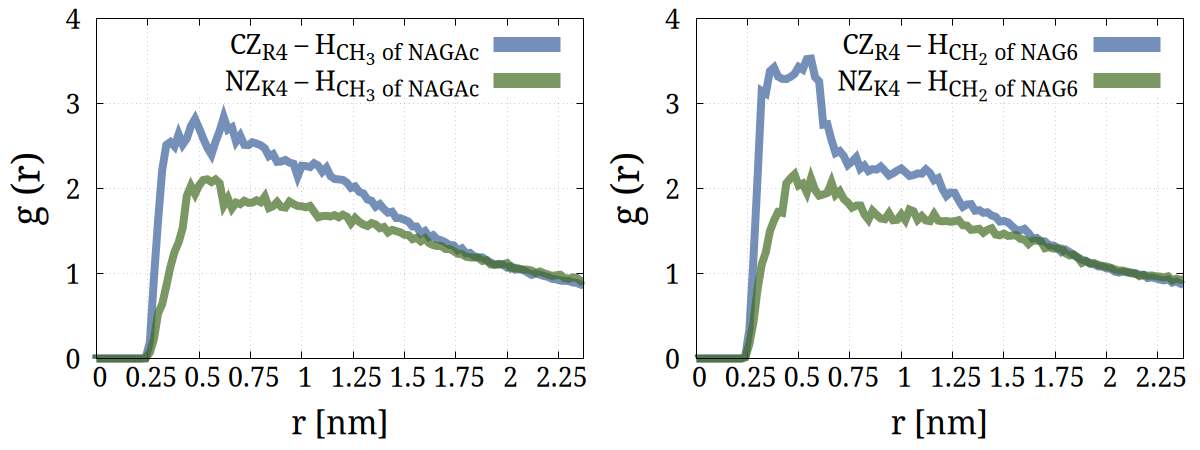}
    \caption{The radial distribution functions for carbon/nitrogen atoms of R4/K4 side-chain cationic groups and methyl/methylene hydrogens of hyaluronan from molecular dynamics simulations with the prosECCo75 model.}
    \label{fgr:MD_RDFs_pi}
\end{figure}
\clearpage

\begin{figure}[htb!]
    \centering
    \includegraphics[width=0.65\textwidth]{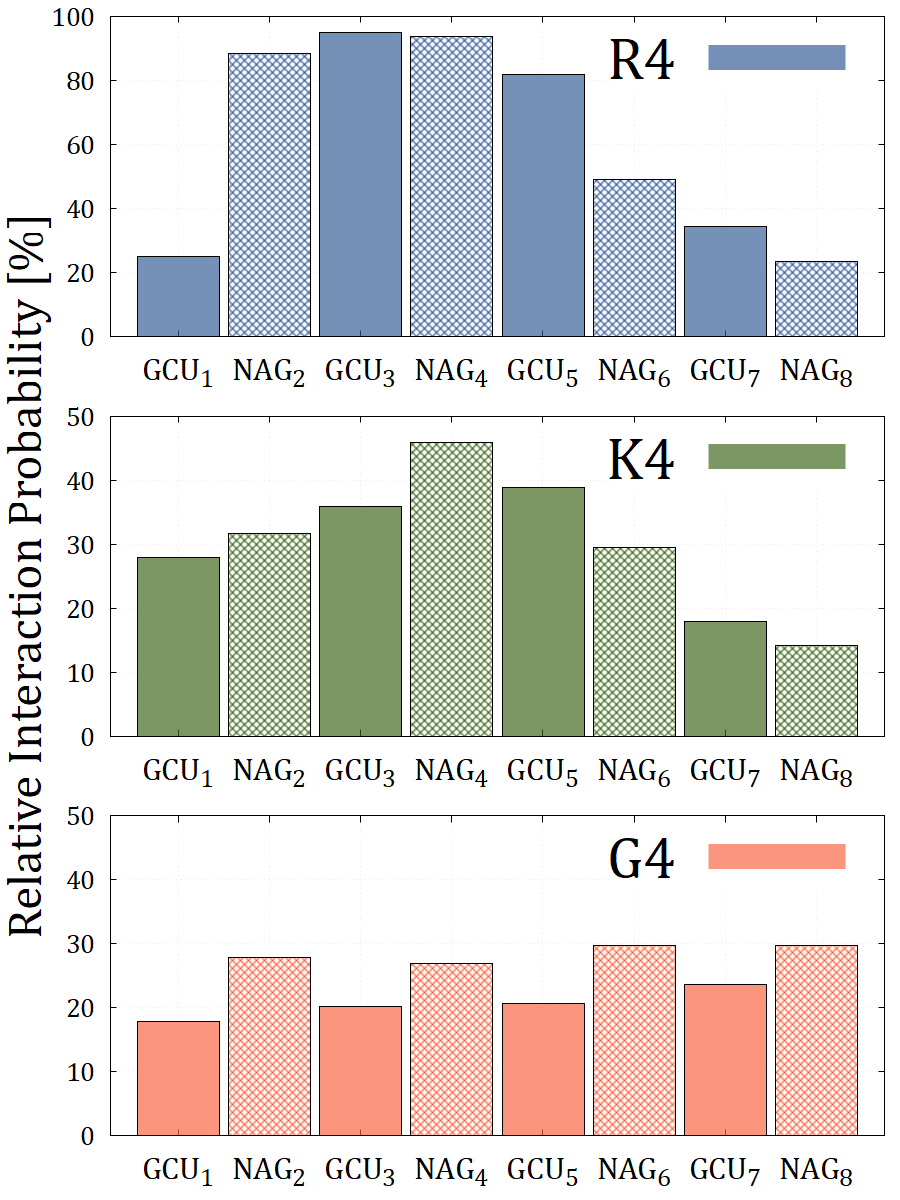}
    \caption{The relative probability of tetrapeptide binding to each of monosaccharides of the hyaluronan octamer as calculated from molecular dynamics simulations with the CHARMM36m model. The relative probability is calculated as the fraction of simulations frames with at least one hydrogen bond between tetrapeptide and monosaccharide with respect to the overall probability of tetrapeptide--hyaluronan binding as given in Figure~\ref{fgr:NMR_MD_summary}d. The probabilites for D-glucuronic acid (GCU monosaccharides) are shown as filled boxes, while the probabilites for N-acetyl-D-glucosamine (NAG monosaccharides) are shown as boxes filled with diagonal crosshatch pattern.}
    \label{fgr:MD_bind_per_sacch_CHARMM}
\end{figure}
\clearpage

\begin{figure}[htb!]
    \centering
    \includegraphics[width=0.7\textwidth]{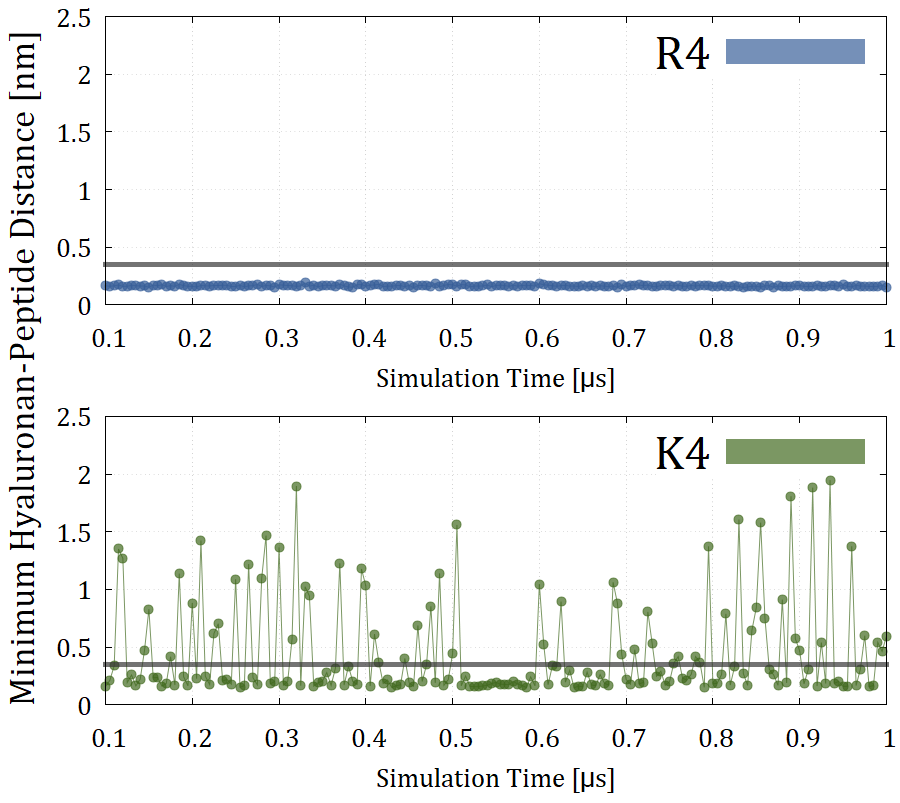}
    \caption{Time dependence of the minimum distance between hyaluronan octamer and tetrapeptides. The data are shown for HA8--R4 and HA8--K4 solutions from molecular dynamics simulations with the CHARMM36m model. The frequency of data points is 5~ns.}
    \label{fgr:MD_lifetime_CHARMM}
\end{figure}
\clearpage

\end{suppinfo}

\end{document}